\documentclass[journal]{IEEEtran}
\usepackage{amsmath,amsfonts,amsthm,amssymb,bm}
\usepackage{algorithmic}
\usepackage{array}
\usepackage[caption=false,font=normalsize,labelfont=sf,textfont=sf]{subfig}
\usepackage{textcomp}
\usepackage{stfloats}
\usepackage{url}
\usepackage{verbatim}
\usepackage{graphicx}
\usepackage{bbding}
\usepackage{multirow}
\usepackage{cite, bbm, xcolor}
\usepackage{refcount}
\usepackage{flafter}
\usepackage{makecell}
\usepackage{diagbox}
\usepackage{booktabs}
\usepackage{algorithm}

\theoremstyle{definition}
\newtheorem{definition}{Definition}
\usepackage[caption=false]{subfig}
\usepackage{float}
\def\BibTeX{{\rm B\kern-.05em{\sc i\kern-.025em b}\kern-.08em
    T\kern-.1667em\lower.7ex\hbox{E}\kern-.125emX}}
\usepackage{balance}
\begin{document}
\title{Joint Beamforming for NOMA Assisted Pinching Antenna Systems (PASS)}

\author{
        Deqiao Gan, Xiaoxia Xu, Jiakuo Zuo, Xiaohu Ge, \textit{Senior Member, IEEE}, and Yuanwei Liu, \textit{Fellow, IEEE}
        \thanks{D. Gan and X. Ge (Corresponding author) are with the School of Electronic Information and Communications, Huazhong University of Science and Technology, Wuhan 430074, Hubei, China. (e-mail: gandeqiao@hust.edu.cn, xhge@mail.hust.edu.cn).}
        \thanks{X. Xu is with the School of Electronic Engineering and Computer Science, Queen Mary University of London, London E1 4NS, U.K. (email: x.xiaoxia@qmul.ac.uk).}
        \thanks{J. Zuo is with the School of Internet of Things, Nanjing University of Post and Telecommunications, Nanjing, Jiangsu, China. (e-mail: zuojiakuo@njupt.edu.cn).}
        \thanks{Y. Liu is with the Department of Electrical and Electronic Engineering, The University of Hong Kong, Hong Kong (e-mail: yuanwei@hku.hk).}
}


\maketitle

\begin{abstract}
Pinching antenna system (PASS) configures the positions of pinching antennas (PAs) along dielectric waveguides to change both large-scale fading and small-scale scattering, which is known as pinching beamforming.
A novel non-orthogonal multiple access (NOMA) assisted PASS framework is proposed for downlink multi-user multiple-input multiple-output (MIMO) communications.
The transmit power minimization problem is formulated to jointly optimize the transmit beamforming, pinching beamforming, and power allocation. To solve this highly nonconvex problem, both gradient-based and swarm-based optimization methods are developed.
1) For gradient-based method, a majorization-minimization and penalty dual decomposition (MM-PDD) algorithm is developed. The Lipschitz gradient surrogate function is constructed based on MM to tackle the nonconvex terms of this problem. Then, the joint optimization problem is decomposed into subproblems that are alternatively optimized based on PDD to obtain stationary closed-form solutions.
2) For swarm-based method, a fast-convergent particle swarm optimization and zero forcing (PSO-ZF) algorithm is proposed. Specifically, the PA position-seeking particles are constructed to explore high-quality pinching beamforming solutions. Moreover, ZF-based transmit beamforming is utilized by each particle for fast fitness function evaluation.
Simulation results demonstrate that: i) The proposed NOMA assisted PASS and algorithms outperforms the conventional NOMA assisted massive antenna system. The proposed framework reduces over 95.22$\%$ transmit power compared to conventional massive MIMO-NOMA systems.
ii) Swarm-based optimization outperforms gradient-based optimization by searching effective solution subspace to avoid stuck in undesirable local optima.
\end{abstract}


\section{Introduction}
\IEEEPARstart{R}{ecent} development of next-generation wireless networks raise high requirements on massive connectivity, ultra-low latency and customizable radio environment \cite{yang2025pinching,kit2021radio}. While conventional massive multiple-input multiple-output (MIMO) systems offer spatial multiplexing gains, their rigid antenna deployment and static beamforming architectures hinder adaptability to dynamic user distributions and line-of-sight (LoS) fluctuations \cite{2013MIMO}. Additionally, it is impossible to reconfigure the fixed antenna array for conventional MIMO. To overcome these limitations, several flexible-antenna techniques, e.g., reconfigurable intelligent surfaces (RISs) \cite{tang2021RIS}, fluid antennas \cite{kit2021fluid, new2024fluid} and movable antennas \cite{ma2024movable, zhu2024movable} have been emerged.

The above mentioned state-of-the-art solutions greatly enhance the controllability of the wireless channel, while they remains confined to the manipulation of small-scale fading and reflection patterns. 
The pinching antenna system (PASS) has arisen as a novel and versatile flexible-antenna system \cite{2022NTTDOCOMO,yuanwei2025pass,wang2025pass,ding2024pass,xu2025pass}, which represents a fundamentally different antenna system design, wherein wireless signals can be flexibly emitted at configurable positions along dielectric waveguides. Specifically, PASS utilizes dielectric waveguides embedded with separately deployed dielectric antennas, referred to as pinching antennas (PAs), that can be flexibly activated and repositioned along the waveguide to form a leaky-wave antenna. Since waveguides can be several meters long, PASS enables continuous spatial reconfiguration of both large-scale path loss and signal phase, thus introducing a novel form of beamforming referred to as pinching beamforming \cite{yuanwei2025pass,wang2025pass}. This pinching beamforming can be enabled by reconfiguring the locations of PAs. The design of PASS was first prototyped by NTT DOCOMO \cite{2022NTTDOCOMO}. And then, it was first introduced into wireless communications by \cite{ding2024pass}. PASS exhibits three critical advantages \cite{yuanwei2025pass,ouyang2025pass,xu2025pass2}:
\textit{1) Adjustable position}: PAs can be easily moved along dielectric waveguides with large adaptable scales for real-time demands.
\textit{2) Path loss configurations and LoS support}: PASS offers a tunable design of antenna system to reconfigure large-scale path loss and builds a strong line-of-sight (LoS) link \cite{ding2025losblockage} to avoid blockage in high-frequency bands.
\textit{3) Flexible beamforming}: By activating PAs at the desired positions, PASS can perform active beam steering or dynamic amplification.

The authors of \cite{ding2024pass} first proposed the PASS design for communication system, which was a low-complexity pinching beamforming architectures for single-user and two-user cases. To increase the gain of PAs, the imapct of LoS blockage on PASS was investigated by \cite{ding2025losblockage}. They analyzed that the LoS blockage served to restrain co-channel interference. The authors of \cite{wang2025pass} developed a downlink PASS architecture with multiple waveguides, each equipped with one PA. They proposed the greedy algorithm to jointly optimize the transmit and pinching beamforming. To further reduce the computational complexity, the physics-based hardware model for PASS was built to reconfigure radiation and simplify signal model. For more generalized scenarios, the downlink PASS architecture that deployed multiple PAs along each waveguide was explored by \cite{xu2025pass}. Moreover, the learning-based method and optimization-based method were analyzed and compared for optimizing the transmit and pinching beamforming. The authors of \cite{xu2025pass2} built an adjustable power radiation model for PASS by configuring power radiation ratios to optimizing transmit and pinching beamforming in general scenarios.

Notably, PAs activated along the same waveguide are vital to transmitting the identical signal, while conventional spatial multiplexing techniques cannot meet the requirements of the case that the number of users surpasses the number of waveguides. 
To overcome this limitation, the development of non-orthogonal multiple access (NOMA) for PASS framework is indispensable, which can serve multiple users sharing the same beam. NOMA has emerged as a promising multiple access technique for next-generation wireless networks due to its ability to support massive connectivity and enhanced spectrum utilization \cite{yuanwei2017NOMA}. More specifically, users in each cluster are based on spatial proximity or channel similarity and are allocated shared resources. This approach naturally aligns with the waveguide-based structure of PASS, where a set of PAs along a single waveguide can serve users in the cluster, enabling both intra-cluster spatial reuse and inter-cluster interference suppression. By enabling multi-user superposition transmission over shared PA configurations, cluster-based NOMA enhances the user-scale multiplexing capability of PASS without proportionally increasing hardware complexity. It also introduces additional flexibility to jointly optimize beamforming and power allocation. The NOMA-assisted PASS was proposed to serve users via downlinks by \cite{ding2025nomapass2}, which can improve the sum rate by the PA matching algorithm. The authors of \cite{ding2025nomapass} proposed a NOMA-aided PASS in two-user case to successively optimize the pinching locations and power allocation by a double-loop successive convex approximation (SCA) algorithm, where the outer loop iteratively updated the power allocation and inner loop updated locations of PAs. This classic scheme is cognitive radio-inspired NOMA in two-user scenario. For generalized scenarios in next generation wireless networks, multi-user access is necessary to be considered to meet real-time demands of users. Moreover, different from conventional orthogonal multiple access (OMA) schemes where each user is assigned a dedicated time/frequency resource \cite{xu2025oma}, NOMA allows multiple users to share the same beam or PA set by superposing their signals in the power domain and separating them via successive interference cancellation (SIC) \cite{yuanwei2017NOMA} at the receivers \cite{wang2025nomapass}.

Although PASS has its unique advantages of reconfigurability and scalability, the current implementation of NOMA for PASS faces two major limitations in support multi-user communication.
\begin{itemize}
    \item \emph{Supprot Multiple NOMA Clusters}: Coventional MIMO-NOMA systems generally utilize cluster-based NOMA design to serve massive user access. However, current NOMA assisted PASS frameworks mainly consider two-user NOMA design and single-waveguide configuration design. How to achieve general cluster-based NOMA in PASS framework still needs to be researched.
    \item \emph{Efficient Optimization Design}: NOMA assisted PASS over multiple clusters require the joint optimization of transmit and pinching beamforming. This leads to a highly coupled optimization problem, which suffers from numerous suboptimal solutions. Hence, efficient optimization techniques relying on both gradient-based and swarm-based methods need to be investigated.
\end{itemize}

To address the above issues, we propose a novel NOMA assisted PASS framework enabled downlink MIMO through a cluster-based design. Specifically, we design a user grouping strategy, which assigns users into different spatial clusters, thus establishing favorable channel effects for each cluster and mitigate both inter-cluster and intra-cluster interference.

A transmit power minimization problem is formulated by jointly optimizing the transmit beamforming, pinching beamforming and power allocation. Further, this paper proposes both gradient-based method and swarm-based method to solve the above joint optimization problem. For gradient-based method, since this minimization problem is highly nonconvex and coupled, we transform it into a tractable form. Then, we design a majorization-minimization and penalty dual decomposition (MM-PDD) algorithm to obtain stationary closed-form solutions. For swarm-based method, we propose a particle swarm optimization and zero forcing (PSO-ZF) algorithm to avoid stuck in undesired local optimum with an acceptable computational complexity. The key contributions of this work are summarized as follows:
\begin{enumerate}
    \item We propose a downlink NOMA assisted PASS framework in MIMO communications, which adjusts the locations of PAs to configure pinching beamforming, thus reducing large-scale path loss and mitigating interference.
    We formulate the transmit power minimization problem to jointly optimize transmit beamforming, and pinching beamforming, and power allocation. To solve this problem, we propose both gradient-based and swarm-based methods.
    \item For gradient-based method, we first convert this highly nonconvex and coupled minimization problem into a tractable problem, and then develop a MM-PDD algorithm. Specifically, the Lipschitz gradient surrogate (LGS) is utilized to combat the nonconvex part of this problem by constructing the upper bound of MM. Based on the PDD, this paper decouples and decomposes the minimization problem into four nested subproblems, and we alternatively optimizes transmit and pinching beamforming and power allocation for the closed-form solutions.
    \item For swarm-based method, we propose a fast-convergent PSO-ZF algorithm with an acceptable computational complexity. To be specific, we construct the position-seeking swarm to intelligently optimize pinching beamforming. Furthermore, we develop a ZF-based transmit beamforming design to obtain the closed-form solutions and avoid stuck in local optimum.
    \item Extensive numerical results verify the effectiveness of the proposed cluster-based NOMA assisted PASS and the developed algorithms, which demonstrates that: i) The proposed NOMA assisted PASS achieves 95.22$\%$ reduction in transmit power compared to conventional massive MIMO-NOMA. ii) The swarm-based method outperforms gradient-based method by searching for effective solution subspace to avoid stuck in local optima.
\end{enumerate}

The remainder of this paper is organized as follows: Section II introduces the system model of NOMA-assisted PASS framework and the problem formulation. Section III addresses the problem by gradient-based optimization method. Section IV presents the swarm-based method to tackle this problem. Numerical results are analyzed in Section V. The conclusion is shown in Section VI.

\textit{Notations}: $\mathbb{C}^{M \times 1}$ denotes the space of $M \times 1$ complex valued vectors, $\text{diag}(\mathbf{x})$ denotes a diagonal matrix whose diagonal elements corresponding to vector $\mathbf{x}$, $\text{blkdiag}(\mathbf{x})$ denotes a block-diagonal matrix. The variable, vector, and matrix are denoted by $x$, $\mathbf{x}$, and $\mathbf{X}$, respectively. $|x|$ denotes the absolute value of a real number and the modulus of a complex number.
The notations $\text{Tr}(\mathbf{X})$ and $\text{rank}(\mathbf{X})$ define the trace and rank of $\mathbf{X}$, respectively.
$I$ is the identity matrix.
$\text{Re}\left\{x\right\}$ and $\text{Im}\left\{x\right\}$ denote the real and image parts of $x$, and $x^{H}$ is the complex conjugate number of $x$. 
$\mathbf{X}^{T}$ and $\mathbf{X}^{H}$ denote the transpose and the Hermitian matrix. 
$\mathcal{CN}(0,\sigma^2)$ represents the distribution of a circularly symmetric complex Gaussian variable (CSCG) with zero mean and $\sigma^2$ variance.

\section{System Model}
\begin{figure}[!t]
    \centering
    \includegraphics[width=3.6in]{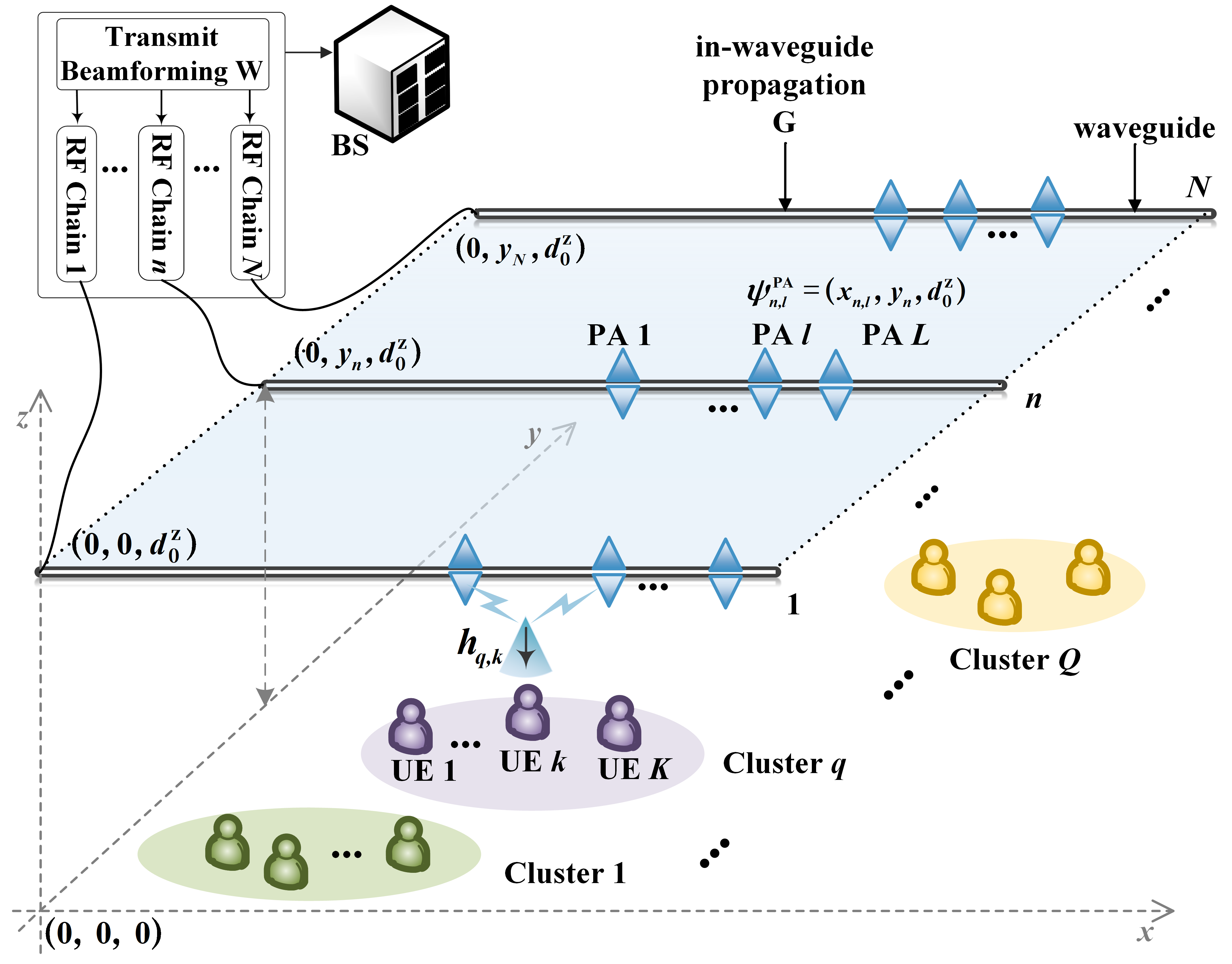}
    \caption{System model: The proposed NOMA assisted PASS framework.}
    \label{systemmodel}
\end{figure}

Consider a NOMA assisted PASS enabled downlink MIMO communication scenario, where a base station (BS) equipped with $n \in \mathcal{N}=\left\{1,...,N\right\}$ dielectric waveguides each pinched along $L$ flexible PAs. The proposed PASS framework exploits cluster-based NOMA, as shown in Fig. \ref{systemmodel}. The distance between each two adjacent waveguides is $d_0$. We assume that waveguides are extended over $x$-axis at the altitude $d^z$ on $z$-axis in an array formed on the $y$-axis with the $d^{\text{y}}_0$ distance between each two adjacent waveguides. The total number of PAs is $M=L\times N$. We denote the location of PA $l \in \mathcal{L}=\left\{1,...,L\right\}$ on waveguide $n$ as ${\bm{\psi}}^{\text{PA}}_{n,l}=[x_{n,l},y_n,d^{\text{z}}_0]$, where $x_{n,l}$ is the tunable pinched location of PA along $x$-axis, $y_n$ is the location of waveguide $n$ along $y$-axis and $y_n=\left(n-1\right){d^{\text{y}}_0}$. The signal is propagated to specific PAs from the feed point of waveguide $n$ with the location ${\bm{\psi}^{\text{PA}}_{n,0}}=[0,y_n,d^{\text{z}}_0]$. Let $\mathbf{x}_{n}=[x_{n,1},x_{n,2},...,x_{n,L}]^T \in \mathbb{R}^{L \times 1}$ represent the x-axis locations of PAs on the waveguide $n$ with $0 \le x_{n,1} < x_{n,2} < x_{n,3} < ... < x_{n,L} \le x^{\max}$, $\forall n \in \mathcal{N}$, where $x^{\max}$ is the maximum lenght of rectangular service area along $x$-axis. And $\mathbf{X}=[\mathbf{x}_{1},\mathbf{x}_{2},...,\mathbf{x}_{N}] \in \mathbb{R}^{L\times N}$ stacks the locations of all PAs. There are $K$ users grouped into $Q$ clusters, denoted by user equipment (UE) set $k \in \mathcal{K}=\left\{ 1,...,K\right\}$ and cluster set $q \in \mathcal{Q}=\left\{ 1,...,Q\right\}$. And the location of UE $k$ in cluster $q$ is denoted as ${\bm{\psi}}^{\text{U}}_{q,k} =[x^{\text{U}}_{q,k},y^{\text{U}}_{q,k},0]$, $\forall k \in \mathcal{K}$, with the same rectangular service area of PAs $0 \le x^{\text{U}}_{q,k} \le x^{\max}$ and $0 \le y^{\text{U}}_{q,k} \le \left(N-1\right){d^{\text{y}}_0}$.

\subsection{PASS Channel Model}

In the MIMO scenario, high-frequency bands, i.e., millimeter-wave (mmWave) and terahertz (THz) \cite{marco2020thz,Elayan2020thz}, are utilized by the NOMA assisted PASS framework. In this paper, we make a practical assumption that the LoS paths between each UE and BS are considered and non-LoS paths are ignored since the non-LoS paths are much weaker than LoS paths \cite{xu2025pass}. Thus, the LoS-dominant model and the geometric free-space spherical model are utilized to formulate the channel vector between PA $l$ along waveguide $n$ and UE $k$ in the cluster $q$,
\begin{equation}
    \label{channelvector}
    h^{H}_{q,k,n,l}=h_{q,k}^{H}(x_{n,l})=\frac{\eta e^{-i{\kappa}\left|{\bm{\psi}}_{q,k}^{\text{U}}-{\bm{\psi}}_{n,l}^{\text{PA}}\right|}}{\left|{\bm{\psi}}_{q,k}^{\text{U}}-{\bm{\psi}}_{n,l}^{\text{PA}}\right|},
\end{equation}
where $\kappa=\frac{2\pi}{\lambda}$ represents the wave-domain number, and $\lambda$ is the wavelength. The constant $\eta=\frac{c}{4\pi f_c}$ depends on the speed of light $c$ and carrier frequency $f_c$. $\frac{\eta}{\left|{\bm{\psi}}_{q,k}^{\text{U}}-{\bm{\psi}}_{n,l}^{\text{PA}}\right|}$ represents the channel gain coefficient. Suppose that UEs in each cluster has the same location along $y$-axis, because the difference in position variation on the $y$-axis is much smaller than that of the $x$-axis. And ${\mathbf{h}}^H_{q,k}(\mathbf{x}_n) \in \mathbb{C}^{1 \times L}$ represents the channel vector from all PAs along waveguide $n$ to UE $k$. Let ${\mathbf{h}}^H_{q,k}(\mathbf{X})=[{{\mathbf{h}}^H_{q,k}\left(\mathbf{x}_{1}\right)},{{\mathbf{h}}^H_{q,k}\left(\mathbf{x}_{2}\right)},\dots,{{\mathbf{h}}^H_{q,k}\left(\mathbf{x}_{N}\right)}] \in \mathbb{C}^{1 \times M}, M=L\times N$ denote the stacked channel vectors from all PAs to UE $k$ in cluster $q$. The element of response vector from the feed point of waveguide $n$ to the corresponding PAs $\mathcal{L}$ can be expressed as
\begin{equation}
    \label{ele_responsevector}
    g_{n,l}(x_{n,l})= \frac{1}{\sqrt{L}}e^{-i\frac{2\pi}{{\lambda}_g}\left|{\bm{\psi}}_{n,0}^{\text{PA}}-{\bm{\psi}}_{n,l}^{\text{PA}}\right|}=\frac{1}{\sqrt{L}}e^{-i\frac{2\pi}{{\lambda}_g}x_{n,l}},
\end{equation}
where $\lambda_g=\frac{\lambda}{n_{\text{eff}}}$ is the guided wavelength, $n_{{\text{eff}}}$ is the effective refractive index of the dielectric waveguide, and $1/{\sqrt{L}}$ represents the power allocation coefficient for all PAs along each waveguide. The above response vector can be expressed as $\mathbf{g}_n(\mathbf{x}_n) \in \mathbb{C}^{L \times 1}$. Then, we can derive the block-diagonal in-wavegudie channel matrix $\mathbf{G}(\mathbf{X}) \in \mathbb{C}^{M \times N}$, which stacks response vectores from all feed points to the corresponding PAs,
\begin{equation}
    \label{blkdiag}
    \begin{aligned}
        \mathbf{G}(\mathbf{X})& =\mathrm{blkdiag}\left(\mathbf{g}_1(\mathbf{x}_1),\mathbf{g}_2(\mathbf{x}_2),\dots,\mathbf{g}_N(\mathbf{x}_N)\right) \\
        & \left.=\left[
        \begin{array}{cccc}
            \mathbf{g}_1\left(\mathbf{x}_1\right) & \mathbf{0} & \ldots & \mathbf{0} \\
            \mathbf{0} & \mathbf{g}_2\left(\mathbf{x}_2\right) & \ldots & \mathbf{0} \\
                \vdots & \vdots & \ddots & \vdots \\
            \mathbf{0} & \mathbf{0} & \ldots & \mathbf{g}_N\left(\mathbf{x}_N\right)
        \end{array}
        \right.\right].
    \end{aligned}
\end{equation}

\subsection{Signal Model}
Since each waveguide is fed by the same signal, which is multiplexed at the baseband by performing transmit beamforming, subsequently converted by the RF chain, and finally fed into the waveguide for radiation. Denote $s_q$ as the transmit signal radiated from the PASS system to all UEs in cluster $q$, which is given by
\begin{equation}
    \label{transmitsignal}
    s_q=\sum_{k=1}^{K}\sqrt{\alpha_{q,k}}s_{q,k},
\end{equation}
where $\alpha_{q,k}$ is the power allocation coefficient for UE $k$ in cluster $q$. Suppose that $s_q \in \mathbb{C}$ is the zero-mean unit-variance stationary process satisfying $\mathbb{E}\left[s_q^H{s_q}\right]=1$. Let the received signal at UE $k$ in cluster $q$ through channels manipulated by PASS system, expressed as
\begin{equation}
    \label{receivedsignal}
    \begin{aligned}
        y_{q,k} = & \underbrace{{\mathbf{h}_{q,k}^H\mathbf{G}}\mathbf{w}_q\sqrt{\alpha_{q,k}}s_{q,k}}_{\text{desired signal}}+\underbrace{\mathbf{h}_{q,k}^H\mathbf{G}\mathbf{w}_q\sum_{i=1,i\neq k}^{K}\sqrt{\alpha_{q,i}}s_{q,i}}_{\text{intra-cluster interference}}\\
        & +\underbrace{\sum_{j=1,j\neq q}^Q\mathbf{h}_{q,k}^H\mathbf{G}\mathbf{w}_j\sum_{i=1}^{K}\sqrt{\alpha_{j,i}}s_{j,i}}_{\text{inter-cluster interference}}+z_k,
    \end{aligned}
\end{equation}
where $z_k \in \mathcal{CN}(0,\sigma^2)$ is the CSCG with zero mean, $\sigma^2$ is the noise power. Moreover, $\mathbf{w}_q \in \mathbb{C}^{N \times 1}$ denotes the transmit beamforming vector for UEs in cluster $q$ and ${\mathbf{h}_{q,k}^H\mathbf{G}}$ represents the corresponding pinching beamforming. The signal-to-interference-and-noise ratio (SINR) of UE $k$ can be formulated as
\begin{equation}
    \label{SINR}
    \begin{aligned}
        & \mathrm{SINR}_{q,{k\to k^{\prime}}} = \\
        & \frac{\left|\mathbf{h}_{q,k^{\prime}}^{H}\mathbf{G}{\mathbf{w}_q}\right|^{2}\alpha_{q,k}}{\left|\mathbf{h}_{q,k^{\prime}}^{H}\mathbf{G}{\mathbf{w}_q}\right|^{2}\!\sum\limits_{i=k+1}^{K}\!\alpha_{q,i}+\!\sum\limits_{j=1,j\neq q}^{Q}\!\left|\mathbf{h}_{q,k^{\prime}}^{H}\mathbf{G}\mathbf{w}_{j}\right|^{2}+\sigma^{2}},\\
        & \forall k,k^{\prime} \in \mathcal{K}, k^{\prime} \ge k,
    \end{aligned}
\end{equation}
where $k^{\prime}=k,k+1,\dots,K$. Then, the corresponding achievable rate is
\begin{equation}
    \label{rate_qk}
    R_{q,k}=\log_2\left(1+\min_{k^{\prime}=k,k+1,\cdots,K}\left\{{\mathrm{SINR}_{q,{k\to k^{\prime}}}}\right\}\right).
\end{equation}

Since the number of UEs in each cluster is no less than 2, each UE needs to eliminate intra-interference by utilizing SIC technology \cite{yuanwei2017NOMA,2022SICli}. To guarantee the SIC performed successfully, the requirements $R_{j,k} \ge R_{q,k}$ with $j \ge k$ should be met according to the principle of NOMA.

\subsection{User Grouping}

\begin{algorithm}[htbp]
    \caption{NOMA Assisted PASS User Grouping Algorithm}
    \label{alg:user_grouping}
    \begin{algorithmic}[1]
    \REQUIRE 
    User set $\mathcal{K} = \{1,2,\dots,K\}$; Number of clusters $Q$; Channel vectors $\{\mathbf{h}_k\}$; Users' positions $\{{\bm{\psi}}_{k}^{\text{U}}\}$; Weighting factor $\varpi$; Parameter $\sigma$.
    \ENSURE 
    User grouping result $\{\mathcal{Q}_1,\mathcal{Q}_2, \dots, \mathcal{Q}_Q\}$.
    \STATE Initialize: Randomly select $M$ users as initial representatives $\{\Xi_q^{(0)}\}$.
    \STATE Set iteration index $t \leftarrow 0$.
    \REPEAT
        \FOR{each user $k =1,2,\dots,K$}
            \FOR{each cluster $q = 1,2,\dots,Q$}
                \STATE Compute similarity 
                $C_{k,\Omega_m^{(t)}}$ by (\ref{jointcorrelation})
            \ENDFOR
            \STATE Assign user $k$ to cluster $q_k^*$ where (\ref{clusterdistribution})
        \ENDFOR
        \FOR{each cluster $q = 1,2,\dots,Q$}
            \STATE Update head user $\Xi_q^{(t+1)}$ as
            $\Xi_q^{(t+1)}$ by (\ref{updatedhead})
        \ENDFOR
        \STATE $t \leftarrow t+1$
    \UNTIL{Convergence by $\{\Xi_q^{(t)}\} = \{\Xi_q^{(t-1)}\}$}
    \RETURN $\{\mathcal{Q}_1,\mathcal{Q}_2, \dots, \mathcal{Q}_Q\}$.
    \end{algorithmic}
    \end{algorithm}

Since the number of users is larger than the number of PAs along each waveguide, it is significant to perform user grouping in NOMA assisted PASS. Based on the equivalent channel correlation between the remaining users, highly channel correlated users are suppose to be assigned into the same cluster, while uncorrelated channel users should be assigned into different clusters to reduce the interference. Moreover, there is not only channel correlation between users, but also location correlation between PAs. Thus, we denote the joint channel and location correlation metric between user $k$ and user $i$ as
\begin{equation}
    \label{jointcorrelation}
    C_{k,i}(\hat{x}_{k,i})=\varpi \frac{{\mathbf{h}^H_k(\hat{x}_{k,i})}{\mathbf{h}_i(\hat{x}_{k,i})}}{\left\|\mathbf{h}_i(\hat{x}_{k,i})\right\|\left\|\mathbf{h}_k(\hat{x}_{k,i})\right\|} + (1-\varpi)e^{-\frac{\left|{\bm{\psi}}_{k}^{\text{U}}-{\bm{\psi}}_{i}^{\text{U}}\right|}{\sigma^2}},
\end{equation}
where $\varpi \in [0,1]$ is the weight that balances channel and space information. Notably, $\mathbf{h}_k$ depends on the estimated location of user $k$, referred as $\hat{x}_{k,i}$. To enable initial user grouping before the joint optimization process, the estimated user location can be employed to compute the approximate channel vectors $\mathbf{h}^H_k(\hat{x}_{k,i})$. These surrogate channel vectors are utilized to calculate the preliminary group matric $C_{k,i}(\hat{x}_{k,i})$. Then, the K-means cluster algorithm \cite{zhu2019usergroup} is utilized to schedule user grouping. Specifically, we select $Q$ users as the head of $Q$ clusters, denoted by $\left\{\Xi_1,\Xi_2,\dots,\Xi_Q\right\}$. According to the joint channel and location correlation (\ref{jointcorrelation}), the remaining users can be distributed to the cluster $q$, measured by
\begin{equation}
    \label{clusterdistribution}
    q^*_k=\underset{k}{\arg} \max_{1 \le q \le Q} C_{k,\Xi_q},
\end{equation}
Then, the head of cluster $q$ is supposed to be updated, which satisfies that its joint channel and location correlation is the lowest with the other clusters. Moreover, the joint channel and location correlation between user $k$ to other users of the other clusters, measured by
\begin{equation}
    \label{correlationcluster}
    \tilde{C}_k = \sum_{1\le j \le K}^{j \notin \mathcal{Q}(k)}C_{k,j},
\end{equation}
where $\mathcal{Q}(k)$ represents the cluster including user $k$. The updated head of cluster $q$ can be expressed as
\begin{equation}
    \label{updatedhead}
    \Xi_q=\underset{j \notin \mathcal{Q}(k)}{\arg} \min \tilde{C}_k.
\end{equation}

Finally, we can obtain the optimal user grouping $\mathcal{Q}=\left\{k|q^*_k=q\right\}, \forall q\in \left\{1,2,\dots,Q\right\}, k \in \left\{1,2,\dots,K\right\}$. The joint channel and location correlation based user grouping algorithm is shown as Algorithm \ref{alg:user_grouping}.

\subsection{Problem Formulation}
In NOMA assisted PASS framework, we aim to minimize the transmit power by jointly optimizing the transmit beamforming $\mathbf{W}=[\mathbf{w}_{1}, \mathbf{w}_{2},\dots,\mathbf{w}_{Q}]$, pinching beamforming $\mathbf{X}=[\mathbf{x}_{1},\mathbf{x}_{2},...,\mathbf{x}_{N}]$, and NOMA power allocation $\bm{\alpha}=[\bm{\alpha}_{1}, \bm{\alpha}_{2},\dots,\bm{\alpha}_{Q}]$. The transmit power minimization problem is formulated as
\begin{subequations}
    \label{min_transmitpower0}
    \begin{align}
            \mathbf{P}_0: \quad & \min_{\mathbf{W}, \mathbf{X}, \bm{\alpha}}\sum_{q=1}^{Q}\left\|\mathbf{w}_{q}\right\|_{2}^{2},
            \label{minw_q}\\
            \mathrm{s.t.~} \quad & R_{q,k} \ge R_{q,k}^{\mathrm{min}},k\in\mathcal{K},q\in\mathcal{Q},
            \label{rateconstraints}\\
            & \sum_{k=1}^{K}\alpha_{q,k}=1,q\in\mathcal{Q}, 
            \label{sumpower}\\
            & 0<\alpha_{q,k}<1,k\in\mathcal{K},q\in\mathcal{Q},
            \label{powerconstraints}\\
            & x_{n,l}-x_{n,l-1} \ge \Delta, l\in\mathcal{L}, n\in\mathcal{N},
            \label{PA_minx}\\
            & x_{n,l}\in[0,x_{n,L}], l\in\mathcal{L}, n\in\mathcal{N},
            \label{PA_x}
    \end{align}
\end{subequations}
constraint (\ref{rateconstraints}) guarantees the minimum data rate of each UE, constraint (\ref{sumpower}) ensures the power allocation coefficient constraint in each cluster, constraint (\ref{powerconstraints}) indicates the variation range of power coefficient for each UE, constraint (\ref{PA_minx}) denotes the minimum antenna space $\Delta$ to avoid mutual coupling, and constraint (\ref{PA_x}) limits the location of each PA to the maximum length of corresponding waveguide.

\vspace{-0.1cm}
\section{Gradient-Based Solution of Transmit Power Minimization For NOMA Assisted PASS}
The transmit power minimization problem is a nonconvex problem, we convert it into the tractable form based on alternating optimization (AO) and augmented Lagrangian (AL) methods \cite{Hong2016aoal} to successively optimize transmit beamforming, pinching beamforming and power allocation, which is tranformed to
\begin{subequations}
    \label{min_transmitpower1}
    \begin{align}
        \mathbf{P}_1: \text{ } &\min_{\substack{\mathbf{W},\mathbf{X},\\ \bm{\alpha},\bm{\beta}}}\sum_{q=1}^{Q}\left\|\mathbf{w}_{q}\right\|_{2}^{2}\! +\!\frac{1}{\mu}\sum_{q=1}^{Q}\sum_{j=1}^{Q}\sum_{k=1}^{K}\left|\mathbf{h}_{q,k}^{H}\mathbf{G}\mathbf{w}_{j}\!-\!\beta_{q,k,j}\right|^{2}\!,
        \label{AOALM}\\
        \mathrm{s.t.~} \text{ }& \frac{\left|\beta_{q,k^{\prime},j}\right|^{2}\alpha_{q,k}}{\left|\beta_{q,k^{\prime},j}\right|^{2}\!\sum\limits_{i=k+1}^{K}\!\alpha_{q,i}\!+\!\sum\limits_{\substack{j=1,\\j\neq q}}^{Q}\!\left|\beta_{q,k^{\prime},j}\right|^{2}\!+\!\sigma^{2}}\ge {\mathrm{SINR}}_{q,k}^{\min},
        \label{SINRconstraint1}\\
        & \text{(\ref{rateconstraints})-(\ref{PA_x})},
        \end{align}
\end{subequations}
where $\beta_{q,k,j}={\mathbf{h}_{q,k}^H\mathbf{G}}\mathbf{w}_j$ is introduced to relax the $\mathbf{P}_1$, ${\mathrm{SINR}}_{q,k}^{\min}$ and ${\mathrm{SINR}}_{q,k}^{\min}$ are the minimum SINR threshold required for data flow of UE $k$ and UE $K$, respectively. Constraint (\ref{SINRconstraint1}) guarantees that the non-last UE can be processed by SIC in cluster $q$, which proves NOMA assisted PASS can decode strong signals. In addition, constraint (\ref{SINRconstraint1}) ensures that the last UE can be processed by SIC in cluster $q$, which means this system can decode its own signal. Meanwhile, these two constraints ensure the successful decoding and user fairness of NOMA assisted PASS.

Due to the strong coupling between the variables, $\mathbf{P}_1$ is a constrained nonconvex optimization problem and belong to nonconvex second-order cone programming (SOCP). To solve this nonconvex problem, we decompose it into three decoupled problems. The transmit beamforming $\mathbf{w}_q$ is is first separated from $\mathbf{P}_1$ to formulate the transmit beamforming optimization problem. The location of PA ${\bm{\psi}}^{\text{PA}}_{n,l}$ and pinching beamforming $\mathbf{h}_{q,k}^H\mathbf{G}$ are further optimized by solving the pinching beamforming optimization problem. Furthermore, The power allocation optimization problem is formulated to optimize the power allocation for UEs $\alpha_{q,k}$. To solve the above subproblems, we adopt the AO and block coordinate descent (BCD) methods and develop a MM-PDD algorithm. Specifically, the MM-PDD algorithm is proposed to refactor the coupling constraints into AL functions, and then successively update these variables during the block updating process, where nonconvex component would be relaxed by majorization minimization \cite{shi2020PDD}.

\subsection{Transmit Beamforming Optimization Problem}
To minimize the transmit power, the transmit beamforming optimization problem can be formulated as
\begin{subequations}
    \label{powerbudget_min}
    \begin{align}
        \mathbf{P}_2: \quad & \min_{\mathbf{w}_{q}}\sum_{q=1}^{Q}\left\|\mathbf{w}_{q}\right\|_{2}^{2} +\frac{1}{\mu}\sum_{q=1}^{Q}\sum_{j=1}^{Q}\sum_{k=1}^{K}\left|\mathbf{h}_{q,k}^{H}\mathbf{G}\mathbf{w}_{j}-\beta_{q,k,j}\right|^{2}
        \label{AL}\\
        \mathrm{s.t.~} \quad & \text{(\ref{rateconstraints})-(\ref{powerconstraints})}.
    \end{align}
\end{subequations}

We introduce two auxiliary variables that the distance and signal phase between PA $l$ and UE $k$ in cluster $q$, denoted as $d^{\text{PU}}_{n,l,q,k}=\sqrt{\left(x_{n,l}-x_{q,k}^{\text{U}}\right)^{2}+\left(y_n-y^{\text{U}}_{q,k}\right)^{2}+{d^{\text{z}}_0}^2}$, $\theta_{n,l,q,k}=\kappa \left(d^{\text{PU}}_{n,l,q,k}+n_{\text{eff}}x_{n,l}\right)$, respectively. The above problem involving nonconvex term is transformed into the following equationally constraint
\begin{equation}
    \label{distanceconstraint}
    b^{\text{U}}_{n,l,q,k} \triangleq d^{\text{PU}}_{n,l,q,k}-\sqrt{\left(x_{n,l}-x_{q,k}^{\text{U}}\right)^{2}+\left(y_n-y^{\text{U}}_{q,k}\right)^{2}+{d^{\text{z}}_0}^2}.
\end{equation}

Define the AL function of $\mathbf{P}_2$ as follows
\begin{equation}
    \label{AL_I2}
    \begin{aligned}
        \mathcal{L}_{w}&\left(\mathbf{w}_{q},\alpha_{q,k},\bm{\psi}_{n,l}^{\mathrm{PA}},\beta_{q,k,j},\lambda^{\text{U}}_{n,l,q,k}\right) = \sum_{q}\left\|\mathbf{w}_{q}\right\|_{2}^{2} \\
        & +\frac{1}{2\rho}\sum_{n,l,q,k}\left|\mathbf{h}_{q,k}^{H}\mathbf{G}\mathbf{w}_{j}-\beta_{q,k,j}+\rho \lambda^{\text{U}}_{n,l,q,k}\right|^{2},
    \end{aligned}
\end{equation}
where $\lambda^{\text{U}}_{n,l,q,k}$ is the dual variable corresponding to (\ref{distanceconstraint}), $\rho$ is the penalty parameter including the $\mu$ of $\mathbf{P}_2$. Let $f_w(\mathbf{w}_q)= \mathcal{L}_{w}\left(\mathbf{w}_{q},\alpha_{q,k},\bm{\psi}_{n,l}^{\mathrm{PA}},\beta_{q,k,j},\lambda^{\text{U}}_{n,l,q,k}\right)$. To distinguish $\mathbf{w}_q$, we introduce a dummy variable $q^{\prime}$ and then derive the derivative of the above AL function 
\begin{equation}
    \label{firstorderL_wq}
    \begin{aligned}
        & \nabla_{\mathbf{w}_{q}}\mathcal{L}_{w}= \nabla f_w(\mathbf{w}_{q})\\
        &= 2\mathbf{w}_{q} + \frac{1}{\rho}\sum_{q^{\prime},k}\mathbf{G}^H\mathbf{h}^*_{q^{\prime},k}\left(\mathbf{h}_{q^{\prime},k}^{H}\mathbf{G}\mathbf{w}_{q}-\beta_{q^{\prime},q,k}+\rho \lambda^{\text{U}}_{q^{\prime},q,k}\right).
    \end{aligned}
\end{equation}

The $\mathcal{L}_{w}\left(\mathbf{w}_{q},\alpha_{q,k},\bm{\psi}_{n,l}^{\mathrm{PA}},\beta_{q,k,j},\lambda^{\text{U}}_{n,l,q,k}\right)$ is a quadratic convex function with respect to $\mathbf{w}_q$ during the BCD updating process. It can be be directly derived to zero and can be quickly analyzed and solved.

\begin{definition}
    The strict analytical solution of AL function (\ref{AL_I2}) can be directly obtained by  derivation to zero of the quadratic convex function on $\mathbf{w}_q$.
    \begin{proof}
        The Lipschitz gradient surrogate (LGS) method \cite{wu2024LGS}, essentially a MM method for approximating strictly convex functions of nonconvex (or convex but complex) functions, is utilized to solve (\ref{AL_I2}). For the above objective function $f_w(\mathbf{w}_q)$, the second derivative (Hessian) matrix is
        \begin{equation}
            \label{secondorderL_wq}
            \nabla^2_{\mathbf{w}_{q}}(\mathbf{w}_q)= \nabla^2 f_w(\mathbf{w}_{q})= 2I + \frac{1}{\rho}\sum_{q^{\prime},k}\mathbf{G}^H\mathbf{h}^*_{q^{\prime},k}\mathbf{h}_{q^{\prime},k}^{H}\mathbf{G}.
        \end{equation}
        
        Since $\nabla^2_{\mathbf{w}_{q}}(\mathbf{w}_q)$ is the constant matrix, the gradient of (\ref{AL_I2}) is Lipschitz-continuous, and hence, the Lipschitz constant can be taken as the spectral norm, i.e., the largest eigenvalue of this second derivative matrix, which can be written as
        \begin{equation}
            \label{eigenvalue_second}
            \varrho^w_{n,l,q,k}=\left\|2I + \frac{1}{\rho}\sum_{q^{\prime},k}\mathbf{G}^H\mathbf{h}^*_{q^{\prime},k}\mathbf{h}_{q^{\prime},k}^{H}\mathbf{G}\right\|_2,
        \end{equation}
        satisfies $\left\|\nabla f_w(\mathbf{w}^{(1)}_q) - \nabla f_w(\mathbf{w}^{(2)}_q)\right\| \le \varrho^w_{n,l,q,k}\left\|\mathbf{w}^{(1)}_q - \mathbf{w}^{(2)}_q\right\|, \forall \mathbf{w}^{(1)}_q,\mathbf{w}^{(2)}_q$. Taking the current iteration point as $\mathbf{w}^{t-1}_q$, the LGS function of $f_w(\mathbf{w}_q)$ is denoted as
        \begin{equation}
            \label{LGS_w}
            \begin{aligned}
                \mathrm{LGS}_{w} = & f_w(\mathbf{w}^{(t-1)}_q) + \text{Re}\left\{\nabla f_w(\mathbf{w}^{(t-1)}_q)^H(\mathbf{w}_q-\mathbf{w}^{(t-1)}_q)\right\} \\ 
                & + \frac{\varrho^w_{n,l,q,k}}{2}\left\|\mathbf{w}_q-\mathbf{w}^{(t-1)}_q\right\|^2,
            \end{aligned}
        \end{equation}
        where $t$ is the number of iterations. The explicit expression of the current gradient is
        \begin{equation}
            \label{gradientf_t1}
            \begin{aligned}
            & \nabla f_w(\mathbf{w}^{(t-1)}_q)= 2\mathbf{w}^{(t-1)}_q + \\
            & \frac{1}{\rho}\sum_{q^{\prime},k}\mathbf{G}^H\mathbf{h}^*_{q^{\prime},k}\left(\mathbf{h}_{q^{\prime},k}^{H}\mathbf{G}{\mathbf{w}_q^{(t-1)}} - \beta_{q^{\prime},q,k} + \rho \lambda^{\text{U}}_{q^{\prime},q,k}\right).
            \end{aligned}
        \end{equation}
        Then, we can get the explicit expression of LGS function (\ref{explicit_LGS_w}).
        \begin{figure*}[t]
            \centering
            \begin{equation}
                \label{explicit_LGS_w}
                \begin{aligned}
                    f_w(\mathbf{w}_q) \le & f_w(\mathbf{w}^{(t-1)}_q) + \frac{\varrho^w_{n,l,q,k}}{2}\left\|\mathbf{w}_q-\mathbf{w}^{(t-1)}_q\right\|^2 + \\
                    & \mathrm{Re}\left\{\left[2\mathbf{w}^{(t-1)}_q + \frac{1}{\rho}\sum_{q^{\prime},k}\mathbf{G}^H\mathbf{h}^*_{q^{\prime},k}\left(\mathbf{h}_{q^{\prime},k}^{H}\mathbf{G}{\mathbf{w}_q^{(t-1)}} - \beta_{q^{\prime},q,k} + \rho \lambda^{\text{U}}_{q^{\prime},q,k}\right)(\mathbf{w}_q-\mathbf{w}^{(t-1)}_q)\right]^H\right\}.
                \end{aligned}
            \end{equation}
        \end{figure*}
        And hence, the updated transmit beamforming vector can be expressed as (\ref{w_q_t}).
        \begin{figure*}[t]
            \centering
            \begin{equation}
                \label{w_q_t}
                    \mathbf{w}_q^{(t)}\! = \!\mathbf{w}_q^{(t-1)}\!-\!\frac{1}{\varrho^w_{n,l,q,k}}\nabla f_w(\mathrm{w}_q^{(t-1)}) \!= \!\mathbf{w}_q^{(t-1)}\!-\!\frac{1}{\varrho^w_{n,l,q,k}} \!\left[2\mathbf{w}^{(t-1)}_q\! +\! \frac{1}{\rho}\sum_{q^{\prime},k}\mathbf{G}^H\!\mathbf{h}^*_{q^{\prime}\!,k}\left(\mathbf{h}_{q^{\prime}\!,k}^{H}\mathbf{G}{\mathbf{w}_q^{(t-1)}}\! - \!\beta_{q^{\prime}\!,q,k} \!+\! \rho \lambda^{\text{U}}_{q^{\prime}\!,q,k}\right)\right]\!.
            \end{equation}
        \end{figure*}
        Finally, the strict analytical solution of AL function can be derived as (\ref{value_wq}), the same as the direct derivation to zero.
        \begin{figure*}
            \begin{equation}
                \label{value_wq}
                    \mathbf{w}^*_q= (\rho I + \sum_{q^{\prime}}^{Q}\sum_{k=1}^{K}\mathbf{G}^H\mathbf{h}^*_{q^{\prime},k}\mathbf{h}_{q^{\prime},k}^{H}\mathbf{G})^{-1} \left[\sum_{q^{\prime}}^{Q}\sum_{k=1}^{K}\mathbf{G}^H\mathbf{h}^*_{q^{\prime},k}\left(\beta_{q^{\prime},q,k}-\rho \lambda^{\text{U}}_{q^{\prime},q,k}\right)\right].
            \end{equation}
        \end{figure*}
    \end{proof}
\end{definition}

\subsection{Pinching Antenna Beamforming Problem}
\subsubsection{Pinching Antenna Position Optimization}
The key idea of AO and BCD is to alternatively fix one subset of variables, either the transmit beamforming or the positions of PAs, and optimize the objective function with respect to the other. With the transmit beamforming $\mathbf{w}_q$, we aim to optimize the position of PAs. The pinching antenna position optimization problen can be formulated as
\begin{subequations}
    \label{min_PAposition}
    \begin{align}
        \mathbf{P}_3: \quad & \min_{\mathbf{X}}\sum_{q=1}^{Q}\sum_{j=1}^{Q}\sum_{k=1}^{K}\left|\mathbf{h}_{q,k}^{H}\mathbf{G}\mathbf{w}_{j}-\beta_{q,k,j}\right|^{2},
        \label{PAposition}\\
        \mathrm{s.t.~} \quad & \text{(\ref{PA_minx})-(\ref{PA_x})}.
        \end{align}
\end{subequations}

Due to the strong coupling of $e^{-i{\kappa}\left|{\bm{\psi}}_{q,k}^{\text{U}}-{\bm{\psi}}_{n,l}^{\text{PA}}\right|}$ and $\frac{\eta}{\left|{\bm{\psi}}_{q,k}^{\text{U}}-{\bm{\psi}}_{n,l}^{\text{PA}}\right|}$, $\mathbf{h}^H_{q,k}$ is highly nonconvex, we utilize the auxiliary variables $d^{\text{PU}}_{n,l,q,k}$, $\theta_{n,l,q,k}$ and corresponding equationally constraints (\ref{distanceconstraint}) and
\begin{equation}
    \label{phaseconstraint}
    b^{\theta}_{n,l,q,k} \triangleq \theta_{n,l,q,k} - \kappa \left(d^{\text{PU}}_{n,l,q,k}+n_{\text{eff}}x_{n,l}\right).
\end{equation}

The AL function of $\mathbf{P}_3$ is denoted as
\begin{equation}
    \label{AL_I3}
    \begin{aligned}
        & \mathcal{L}_{\psi}\left(x_{n,l}\right) = \sum_{q,k,j}\left|\mathbf{h}_{q,k}^{H}\mathbf{G}\mathbf{w}_{j}-\beta_{q,k,j}\right|^{2} + \\
        & \frac{1}{2\rho}\sum_{n,l,q,k}\left(\left|b^{\text{U}}_{n,l,q,k}+\rho \lambda^{\text{U}}_{n,l,q,k}\right|^{2} + \left|b^{\theta}_{n,l,q,k}+\rho \lambda^{\theta}_{n,l,q,k}\right|^2\right),
    \end{aligned}
\end{equation}
where $\lambda^{\theta}_{n,l,q,k}$ is the dual variable corresponding to this constraint and $\rho$ is updated based on $\mathbf{P}_2$. Let $f_c(x_{n,l})=\sum_{q,k,j}\left|\mathbf{h}_{q,k}^{H}\mathbf{G}\mathbf{w}_{j}-\beta_{q,k,j}\right|^{2}$ represent the channel error term, which depends on $d^{\text{PU}}_{n,l,q,k}$. We can simplify the $d^{\text{PU}}_{n,l,q,k}=\sqrt{\left(x_{n,l}-x_{q,k}^{\text{U}}\right)^{2}+C_{q}}$ with the constant $C_q=\left(y_n-y^{\text{U}}_{q,k}\right)^{2}+{d^{\text{z}}_0}^2$ in each cluster. And the critical non-convex term has been replaced by
\begin{equation}
    \label{distanceconst_rep}
    \begin{aligned}
        & b^{\text{U}}_{n,l,q,k} \approx  d^{\text{PU}}_{n,l,q,k} - \\
        & \left[{d^{\text{PU}}_{n,l,q,k}}^{(t-1)} + \frac{\left(x^{(t-1)}_{n,l}-x_{q,k}^{\text{U}}\right)\left(x_{n,l}-x_{n,l}^{(t-1)}\right)}{{d^{\text{PU}}_{n,l,q,k}}^{(t-1)}}\right].
    \end{aligned}
\end{equation}

We can obtain the second derivative matrix of $f_c$ at iteration point $\mathbf{x}_n^{(t-1)}$ by Taylor expansion, that is $\nabla^2_{\psi}f_c = \sum_{q,k,j}2\text{Re}\left\{\nabla_{\mathbf{x}_n} \tilde{\beta}_{q,k,j}(\mathbf{x}_n^{(t-1)})\left[\nabla_{\mathbf{x}_n} \tilde{\beta}_{q,k,j}(\mathbf{x}_n^{(t-1)})\right]^H \right\}$. From (\ref{distanceconstraint}) and (\ref{phaseconstraint}), the derivatives of two corresponding non-convex terms are derived as $\nabla _{\psi}{b^{\text{U}}_{n,l,q,k}} = -\frac{(x_{n,l}-x^{\text{U}}_{q,k})}{d^{\text{PU}}_{n,l,q,k}(\mathbf{x}_n)}e_{n,l}$ and $\nabla _{\psi}{b^{\text{U}}_{n,l,q,k}} = -\kappa \frac{(x_{n,l}-x^{\text{U}}_{q,k})}{d^{\text{PU}}_{n,l,q,k}(\mathbf{x}_n)}e_{n,l}$, where $e_{n,l}$ is $NL$-dimensional unit vector.
And the second derivative matrices of them are expressed as $\nabla^2_{\psi}b^{\text{U}}_{n,l,q,k}=-\frac{C_q}{\left[d^{\text{PU}}_{n,l,q,k}(\mathbf{x}_n)\right]^3}e_{n,l}e_{n,l}^T$ and $\nabla^2_{\psi}b^{\theta}_{n,l,q,k}=-\kappa \frac{C_q}{\left[d^{\text{PU}}_{n,l,q,k}(\mathbf{x}_n)\right]^3}e_{n,l}e_{n,l}^T$, respectively.
Then, the second derivative of distance constraint in (\ref{AL_I3}) can be derived as 
\begin{equation}
    \label{secondderivative_I3}
    \begin{aligned}
        \nabla^2_{\psi}{\mathcal{L}_{\psi}} = & \frac{1}{\rho}\sum_{n,l,q,k}\nabla_{\psi}b^{\text{U}}_{n,l,q,k}\nabla_{\psi}({b^{\text{U}}_{n,l,q,k}})^T \\
        & +\frac{1}{\rho}\sum_{n,l,q,k}(b^{\text{U}}_{n,l,q,k}+\rho \lambda^{\text{U}}_{n,l,q,k})\nabla^2_{\psi}b^{\text{U}}_{n,l,q,k},
    \end{aligned}
\end{equation}
and the second derivative of phase constraint in (\ref{AL_I3}) can be formulated as
\begin{equation}
    \label{secondderivative_theta}
    \begin{aligned}
        \nabla^2_{\psi}b^{\theta}_{n,l,q,k} =& \frac{1}{\rho}\sum_{n,l,q,k}\nabla_{\psi}b^{\theta}_{n,l,q,k}\nabla_{\psi}({b^{\theta}_{n,l,q,k}})^T\\
        & + \frac{1}{\rho}\sum_{n,l,q,k}(b^{\theta}_{n,l,q,k}+\rho \lambda^{\theta}_{n,l,q,k})\nabla^2_{\psi}b^{\theta}_{n,l,q,k}.
    \end{aligned}
\end{equation}

Thus, the entire second derivative matrix can be formulated as (\ref{secondorderLpsi}).
\begin{figure*}[ht]
    \begin{equation}
        \label{secondorderLpsi}
        \begin{aligned}
            \nabla^2_{\psi}\mathcal{L}_{\psi}(\text{x}_n) = & \sum_{q,k,j}2\text{Re}\left\{\nabla_{\mathbf{x}_n} \tilde{\beta}_{q,k,j}(\mathbf{x}_n^{(t-1)})\left[\nabla_{\mathbf{x}_n} \tilde{\beta}_{q,k,j}(\mathbf{x}_n^{(t-1)})\right]^H \right\} + \\
            & \frac{1}{\rho}\sum_{n,l,q,k}\left[\nabla_{\psi}b^{\text{U}}_{n,l,q,k}\nabla_{\psi}({b^{\text{U}}_{n,l,q,k}})^T+(b^{\text{U}}_{n,l,q,k}+\rho \lambda^{\text{U}}_{n,l,q,k})\nabla^2_{\psi}b^{\text{U}}_{n,l,q,k}\right] + \\
            & \frac{1}{\rho}\sum_{n,l,q,k}\left[\nabla_{\psi}b^{\theta}_{n,l,q,k}\nabla_{\psi}({b^{\theta}_{n,l,q,k}})^T+(b^{\theta}_{n,l,q,k}+\rho \lambda^{\theta}_{n,l,q,k})\nabla^2_{\psi}b^{\theta}_{n,l,q,k}\right].
        \end{aligned}
    \end{equation}
\end{figure*}
Furthermore, the pinching antenna position optimization $\mathbf{P}_3$ can be updated as the convex quadratic objective optimization, which can be expressed as
\begin{subequations}
    \label{convexquad_I3}
    \begin{align}
        \nonumber
        & \min_{\psi_{n,l}^{\mathrm{PA}}} \text{ } [\frac{1}{2}(\mathbf{x}_n\!-\!\mathbf{x}_n^{(t-1)})^T\nabla^2_{\psi}\mathcal{L}_{\psi}(\mathbf{x}_n\!-\!\mathbf{x}_n^{(t-1)})\!+\!\mathcal{L}_{\psi}(\mathbf{x}_n^{(t-1)}) \\
        & \qquad +\nabla_{\psi}\mathcal{L}_{\psi}(\mathbf{x}_n^{(t-1)})^T(\mathbf{x}_n-\mathbf{x}_n^{(t-1)})]
        \label{convexobjective_I3}\\
        \mathrm{s.t.~} & \text{(\ref{PA_minx})-(\ref{PA_x})},       
    \end{align}
\end{subequations}
where gradient of (\ref{AL_I3}) can be expressed as (\ref{gradient_I3}),
\begin{figure*}[ht]
    \begin{equation}
        \label{gradient_I3}
        \begin{aligned}
            \nabla_{\psi}\mathcal{L}_{\psi}(\mathbf{x}_n^{(t-1)})= & \nabla_{\psi}f_c(\mathbf{x}_n^{(t-1)}) + \frac{1}{\rho}\sum_{n,l,q,k}(b^{\text{U}}_{n,l,q,k} +  \rho \lambda^{\text{U}}_{n,l,q,k})\nabla^2_{\psi}b^{\text{U}}_{n,l,q,k}(\mathbf{x}_n^{(t-1)}) \\ 
            & + \frac{1}{\rho}\sum_{n,l,q,k}(b^{\theta}_{n,l,q,k} + \rho \lambda^{\theta}_{n,l,q,k})\nabla^2_{\psi}b^{\theta}_{n,l,q,k}(\mathbf{x}_n^{(t-1)}).
        \end{aligned}
    \end{equation}
\end{figure*}
and constant term is
\begin{equation}
    \label{constant_I3}
    \begin{aligned}
        \mathcal{L}_{\psi}(\mathbf{x}_n^{(t-1)}) = & f_c(\mathbf{x}_n^{(t-1)}) + \\
        & \frac{2}{\rho}\sum_{n,l,q,k}\left(b^{\text{U}}_{n,l,q,k}(\mathbf{x}_n^{(t-1)}) + \rho \lambda^{\text{U}}_{n,l,q,k}\right)^2 +\\
        & \frac{2}{\rho}\sum_{n,l,q,k}\left(b^{\theta}_{n,l,q,k}(\mathbf{x}_n^{(t-1)}) + \rho \lambda^{\theta}_{n,l,q,k}\right)^2,
    \end{aligned}
\end{equation}
where the Lipschitz-constant can be derived as $2\kappa^2 n^2_{\text{eff}} \left|\frac{\eta {(\mathbf{G}\mathbf{w}_j)}^H\beta_{q,k,j}}{d^{\text{PU}}_{n,l,q,k}}\right|$.
The pinching antenna position optimization can be transformed into
\begin{equation}
    \label{x_n}
    \mathbf{x}_n^{(t)}= \arg \min_{\mathbf{x}_n} \nabla^2_{\psi}\mathcal{L}_{\psi}(\text{x}_n),
\end{equation}
during the iteration process, we can obtain
\begin{equation}
    \label{optimal_xn}
        \mathbf{x}^*_n = \mathbf{x}^{(t-1)}_n + \frac{\text{Im}\left\{\frac{\eta {(\mathbf{G}\mathbf{w}_j)}^H\beta_{q,k,j}}{d^{\text{PU}}_{n,l,q,k}} e^{i\kappa n_{\text{eff}}\mathbf{x}_n^{(t-1)}}\right\}}{\kappa n_{\text{eff}}\left|\eta {(\mathbf{G}\mathbf{w}_j)}^H\beta_{q,k,j}\right|},
\end{equation}
then, the LGS-updated explicit expression optimal $d^{\text{PU}*}_{n,l,q,k}$ is expressed as
\begin{equation}
    \label{optimal_dPU}
        d^{\text{PU}*}_{n,l,q,k} = \frac{1}{\kappa}\theta^*_{n,l,q,k} - n_{\text{eff}}x^*_{n,l},
\end{equation}
and optimal $\theta^*_{n,l,q,k}$ is derived as (\ref{optimal_theta}).
\begin{figure*}[t]
    \begin{equation}
        \label{optimal_theta}
        \theta^*_{n,l,q,k}=\frac{\frac{1}{\rho} \kappa({d^{\text{PU}}_{n,l,q,k}}^{(t-1)}+n_{\mathrm{eff}}x_{n,l})+\frac{\eta}{\sqrt{L}}|\lambda^{\theta}_{n,l,q,k} + \frac{\eta e^{-i\theta^{(t-1)}_{n,l,q,k}}}{\sqrt{L}\rho}| \theta_{n,l,q,k}^{(t-1)}-\frac{\eta}{\sqrt{L}} \text{Im}\left\{\lambda^{\theta}_{n,l,q,k} + \frac{\eta e^{-i\theta^{(t-1)}_{n,l,q,k}}}{\sqrt{L}\rho}e^{j\theta_{n,l,q,k}^{(t-1)}}\right\}}{\frac{1}{\rho}+\frac{\eta}{\sqrt{L}}|\lambda^{\theta}_{n,l,q,k} + \frac{\eta e^{-i\theta^{(t-1)}_{n,l,q,k}}}{\sqrt{L}\rho}|}.
    \end{equation}
\end{figure*}

The interior point method \cite{nem2008interior} is adopted to obtain optimal solution of pinching antenna position ${\bm{\psi}}^{\text{PA}}_{n,l}$.

\subsubsection{Pinching Beamforming Optimization}
With the optimal PA location, the pinching beamforming vector can be optimizad by the same minimization problem $\mathbf{P}_3$. We introduce the auxiliary variable $\mathbf{u}_{q,k}=\mathbf{h}_{q,k}^H\mathbf{G} \in \mathbb{C}^{1\times N}$ to represent the pinching beamforming vector, where $\mathbf{u}_{q,k}=[\mathbf{u}_{1,q,k},\mathbf{u}_{2,q,k},\dots,\mathbf{u}_{N,q,k}] \in \mathbb{C}^{1\times N}$, $\mathbf{u}_{n,q,k}=[u_{n,1,q,k},u_{n,2,q,k},\dots,u_{n,L,q,k}]$, $u_{n,l,q,k}=\frac{\eta e^{-i\theta_{n,l,q,k}}}{\sqrt{L}d^{\text{PU}}_{n,l,q,k}}$ denotes the beamforming coefficient of UE $k$ in cluster $q$ radiated by the PA $l$. We stack the directional vectors of all UE $\left\{\mathbf{u}_{q,k}\right\}$ to form a matrix $\mathbf{U}\in \mathbb{C}^{K\times M}$. And $\beta_{q,k,j}=\mathbf{u}^H_{q,k}\mathbf{w}_j$ can be obtained. From $\mathbf{P}_3$ (\ref{PAposition}), the objective function can be transformed into 
\begin{equation}
    \label{min_u}
    \min_{\mathbf{u}_{q,k}}\left|\mathbf{u}_{q,k}\mathbf{W}-\bm{\beta}_{q,k}\right|^2,
\end{equation}
where $\mathbf{W}=[\mathbf{w}_1,\mathbf{w}_2,\dots,\mathbf{w}_Q] \in \mathbb{C}^{N \times Q}$ and $\bm{\beta}_{q,k}=[\beta_{q,k,1},\beta_{q,k,2},\dots,\beta_{q,k,Q}]^T \in \mathbb{C}^{Q \times 1}$. This (\ref{min_u}) is a complex least squares problem, it can be solved by the classical complex least squares method \cite{mark2007square}, the closed-form solution can be derived as
\begin{equation}
    \label{optimal_u}
    \mathbf{u}^*_{q,k}=(\mathbf{W}\mathbf{W}^H)^{-1}\mathbf{W}\bm{\beta}_{q,k},
\end{equation}
\begin{equation}
    \label{optimal_beta}
    \beta^*_{q,k,j}=\left[(\mathbf{W}\mathbf{W}^H)^{-1}\mathbf{W}\bm{\beta}_{q,k}\right]^H\mathbf{w}^*_q.
\end{equation}

\subsection{Power Allocation Optimization Problem}
In the NOMA assisted PASS framework, UEs within the same cluster decode each other through SIC, and there is the mutual interference between different clusters. This would result in less efficient power allocation. To ensure the minimum required rate of each UE, the power allocation optimization problem can be formulated as
\begin{subequations}
    \label{min_powerallocation}
    \begin{align}
        \mathbf{P}_4: \quad & \min_{\alpha_{q,k}, \beta_{q,k,j}}\sum_{q=1}^{Q}\sum_{j=1}^{Q}\sum_{k=1}^{K}\left|\mathbf{h}_{q,k}^{H}\mathbf{G}\mathbf{w}_{j}-\beta_{q,k,j}\right|^{2},
        \label{alpha}\\
        \mathrm{s.t.~} \quad & \text{(\ref{sumpower})-(\ref{PA_x}),(\ref{SINRconstraint1})},
        \end{align}
\end{subequations}
where (\ref{SINRconstraint1}) is nonlinear, recursive inequality constraint that expresses the minimum power allocation requirements. And these constraints can be transformed into
\begin{equation}
    \label{uniform_powercons}
    \begin{aligned}
        & \left|\beta_{q,k^{\prime},q}\right|^2\!\left(\!\alpha_{q,k}\!-\!{\mathrm{SINR}}_{q,k}^{\min}\!\sum_{i=k+1}^{K}\!\alpha_{m,i}\!\right)\! \ge \! \\
        & {\mathrm{SINR}}_{q,k}^{\min}\!\left(\sum_{j=1,j\neq q}^Q\!\left|\beta_{q,k^{\prime},j}\right|^2\!+\!\sigma^2\!\right),
    \end{aligned}
\end{equation}
where $\left|\beta_{q,k^{\prime},q}\right|^2$ and $\left|\beta_{q,k^{\prime},j}\right|^2$ represent channel projection in the cluster $q$ and inter-cluster interference, respectively. It can be obtained 
\begin{equation}
    \label{alpha_lowbound}
    \begin{aligned}
        \alpha_{q,k} \ge & \frac{{\mathrm{SINR}}_{q,k}^{\min}\left(\sum_{j=1,j\neq q}^Q\left|\beta_{q,k^{\prime},j}\right|^2+\sigma^2\right)}{\left|\beta_{q,k^{\prime},q}\right|^2} \\
        & + {\mathrm{SINR}}_{q,k}^{\min}\sum_{i=k+1}^{K}\alpha_{m,i}.
    \end{aligned}
\end{equation}

Let $a_k \triangleq \frac{{\mathrm{SINR}}_{q,k}^{\min}\left(\sum_{j=1,j\neq q}^Q\left|\beta_{q,k^{\prime},j}\right|^2+\sigma^2\right)}{\left|\beta_{q,k^{\prime},q}\right|^2}$. Then, the recursive constraint becomes $\alpha_{q,k} \ge a_{q,k} + {\mathrm{SINR}}_{q,k}^{\min}\sum_{i=k+1}^{K}\alpha_{m,i}$. We derive the closed-form solution utilizing a backward iterative approach of dynamic programming
\begin{equation}
    \label{non-normalizedform}
    \tilde{\alpha}_{q,k} = \sum_{i=k}^{K}(\prod_{p=k}^{i-1}c_p)a_{q,i},
\end{equation}
where $\prod_{p=k}^{i-1}c_p=1$. The normalized closed-form power allocation can be derived as
\begin{equation}
    \label{normalizedform_alpha}
    \alpha^*_{q,k}=\frac{\sum_{i=k}^{K}(\prod_{p=k}^{i-1}c_p)a_{q,i}}{\sum_{v=1}^{K}\sum_{i=v}^{K}(\prod_{p=v}^{i-1}c_p)a_{q,i}}.
\end{equation}

\begin{algorithm}[!ht]
    \caption{Iterative MM-PDD Algorithm for Joint Transmit And Pinching Beamforming And Power Allocation Optimization}
    \label{MMPDD}
    \begin{algorithmic}[1]
        \REQUIRE $(x^{\text{U}}_{n,l},y^{\text{U}}_{n,l})$, $\kappa$, ${\mathrm{SINR}}_{q,k}^{\min}$. 
    \STATE \textbf{Initialize:} $\mathbf{w}_q^{(0)}$, $\theta_{n,l,q,k}^{(0)}$, $\beta_{q,k,j}^{(0)}$, $\alpha_{q,k}^{(0)}$, $\mathbf{u}_{q,k}^{(0)}$, $\rho > 0$, $t \gets 0$
    \WHILE {$t < T$}
      \STATE \hspace{-2mm} {\textbf{STEP A: Transmit Beamforming Optimization}}
    
      \FOR{each cluster $q = 1$ to $Q$}
        \STATE Solve 
          $\displaystyle \mathbf{P}_2$ (\ref{AL})
        \STATE Update
          $\mathbf{w}_q^{(t+1)}$ as Closed-Form Solution by LGS
      \ENDFOR
    
      \STATE \hspace{-2mm} {\textbf{STEP B: Pinching Beamforming Optimization}}
    
      \FOR{each PA with $x_{n,l}$ and UE $k$ in cluster $q$}
      \STATE Solve: $\displaystyle \mathbf{P}_3$ (\Ref{PAposition})
        \STATE Update phase and distance
          $\theta_{n,l,q,k}^{(t+1)}, d_{n,l,q,k}^{\text{PU}(t+1)}$ via Constructing Local Quadratic Upper Bounds by LGS
        \STATE Update position
          $\mathbf{x}_n, x_{n,l}^{(t+1)} $ by Interior Point Method
      \STATE Solve
          $\displaystyle$ (\ref{min_u})
        \STATE Update
          $\mathbf{u}_{q,k}^{(t+1)}, \beta^{(t+1)}_{q,k,j}$ as Closed-Form Solution by Complex Least Squares Method
      \ENDFOR
    
      \STATE \hspace{-2mm} {\textbf{STEP C: Power Allocation Optimization}}
    
      \FOR{each UE $k$}
        \FOR{each cluster $j$}
          \STATE Update
            $\beta_{q,k,j}^{(t+1)}$ by $\mathbf{h}_{q,k}^H G \mathbf{w}_j^{(t+1)}$
        \ENDFOR
    
        \STATE Solve
          $\displaystyle \mathbf{P}_4$ (\ref{alpha}) subject to (\ref{uniform_powercons})
          \STATE Update
            $\alpha_{q,k}^{(t+1)}$ as Closed-Form Solution by Backward Iterative Approach
      \ENDFOR
    
      \STATE $t \gets t+1$
    \ENDWHILE
    \ENSURE $\mathbf{w}_q$, $\mathbf{x}_n$, $d^{\text{PU}}_{n,l,q,k}$, $\theta_{n,l,q,k}$, $\mathbf{u}_{q,k}$, $\beta_{q,k,j}$, $\alpha_{q,k}$
    \end{algorithmic}
\end{algorithm}

Based on the above methods, the joint transmit and pinching beamforming and power allocation optimization algorithm in the NOMA assisted PASS is shown in Algorithm \ref{MMPDD}.

\subsection{Computational Complexity Analysis}

The computational complexity of each subproblem in the proposed iterative MM-PDD algorithm in the NOMA assisted PASS framework is analyzed.
Subproblem $\mathbf{P}_2$ updates the transmit beamforming vector $\mathbf{w}_q$ using a LGS solution with complexity $\mathcal{O}(QM)$. Subproblem $\mathbf{P}_3$ optimizes the PA location, phase variables and pinching beamforming vectors $\mathbf{u}_{q,k}$ via LGS, yielding $\mathcal{O}(Q M K x^{\max})$. Subproblem $\mathbf{P}_4$ performs power allocation, either via a recursive closed-form expression, resulting in a worst-case complexity of $\mathcal{O}(K^2 Q)$. Therefore, the overall computational complexity is $\mathcal{O}\left( Q M + M Q K d + K^2 Q \right)$.
\section{Swarm-Based Solution of Transmit Power Minimization For NOMA Assisted PASS}
Although the above gradient descent-based method can solve the $\mathbf{P}_2$, there are still two challenges. On the one hand, $\mathbf{P}_1$ has a large number of local optimal. Since PA position $\mathbf{X}$ directly affect the free-space channel gains, the phase shifts and the in-waveguide channel matrix, the objective functions, e.g., transmit power, are highly nonlinear and nonconvex with respect to $\mathbf{X}$ according to constraints (\ref{rateconstraints})-(\ref{PA_x}). On the other hand, due to the wide deployment range of PAs, there is significant oscillation in space path loss, and then results in the large performance gap between local optima. In this section, to address the above challenges, we propose the particle swarm optimization method with its global search capability, gradient-free dependence, parallelizability and constraint adaptability to search for the optimal positions of PAs. Furthermore, the ZF-based method is proposed to optimize the transmit beamforming and pinching beamforming and power allocation.

\subsection{Particle Swarm Optimization For Pinching Antenna Position Optimization}
To ensure the communication quality of each user while reducing transmission costs, we aim to minimize the total transmit power. Thus, the total transmit power minimization problem is formulated as $\mathbf{P}_0$ (\ref{min_transmitpower0}). Denote that each waveguide is a swarm, and each swarm has $L$ particles. Let the position-seeking particles in each swarm as $\mathbf{p}_n=[p_{n,1},p_{n,2},\dots,p_{n,L}]^T$, which stacks particles of all PAs along waveguide $n$ \cite{zhu2019usergroup,jiang2025pso}. During the $t$-th iterations, the tuple of particle $p^{(t)}_{n,l}$ is denoted as
\begin{equation}
    \label{particle_p}
    p^{(t)}_{n,l} \triangleq \left\{x^{(t)}_{n,l},v_{n,l}^{(t)},\tilde{x}^{(t)}_{n,l}\right\},
\end{equation}
where $x^{(t)}_{n,l}$ and $\tilde{x}^{(t)}_{n,l}$ represent the current position and recorded best position of PA $l$ along waveguide $n$, $v_{n,l}^{(t)}$ is the update velocity of this particle. In the $(L \times N)$-dimensional searching space, the particles matrix $\mathbf{P}=[\mathbf{p}_1,\mathbf{p}_2,\dots,\mathbf{p}_N]$ in all swarms are randomly initialized with PA position matrix $\mathbf{X} \in \mathbb{R}^{L \times N}$ and corresponding velocity matrix $\mathbf{V} \in \mathbb{R}^{L \times N}$, which consists of all particles' update velocities $v_{n,l}$, denoted as
\begin{equation}
    \label{velocity_p}
    v_{n,l}^{(t+1)} = a_0 v_{n,l}^{(t)} + a_1 r_1 (J_{n,l}-x^{(t)}_{n,l}) + a_2 r_2 (O_{n,l}-x^{(t)}_{n,l}),
\end{equation}
where $J_{n,l}$ denotes the local best position for particle $p_{n,l}$ and $O_{n,l}$ represents the global best position for $p_{n,l}$, $a_0$ is the inertia weight that balances seeking, $a_1$ and $a_2$ are cognitive and social ratios, $r_1$ and $r_2$ are generated randomly with the uniform distribution $[0,1]$, respectively. Without loss of generality, $a_0$ decreases linearly from maximum to minimum during the iterations. The inner-swarm exchange depends on global best position $O_{n,l}$, which is evaluated by the fitness function $f_0{(x_{n,l})}=\sum_{Q} \| \mathbf{w}_q \|^2$. For each iteration, the particles out of
the boundaries are adjusted to the boundary range $[0, x^{\max}]$. Thus, the updated position of PA $l$ is expressed as
\begin{equation}
    \label{position_p}
    x^{(t+1)}_{n,l} = x^{(t)}_{n,l} + v_{n,l}^{(t)}.
\end{equation}


\vspace{-0.2cm}
\subsection{Zero Forcing-Based Solution of Joint Transmit And Pinching Beamforming Optimization}

Searching methods for $\mathbf{P}_0$ minimization would result in high complexity, since it must compute matrix inversion at each iteration. With the optimal positions of PAs derived in $\text{Part A}$ of this section, ZF can effectively reduce the computational complexity \cite{wang2025pass}. In this section, we propose a joint transmit and pinching beamforming optimization algorithm based on ZF to minimize the transmit power. This algorithm can eliminate the inter-cluster interference by ZF and intra-cluster interference of weakest user by SIC to decrease the optimization complexity. According to the optimal global position matrix $\mathbf{X}^*$ and equivalent channel matrix $\mathbf{U} \in \mathbb{C}^{K \times M}$, we can derive the ZF transmit beamforming matrix with any given pinching beamforming matrix, expressed as
\begin{equation}
    \label{ZFtransmitbeam}
    \mathbf{W}=\mathbf{U}(\mathbf{X}^*){(\mathbf{U}^H(\mathbf{X}^*)\mathbf{U}(\mathbf{X}^*))}^{-1}{\sqrt{\mathbf{P}}},
\end{equation}
where the power matrix to UEs in the proposed NOMA assisted PASS is denoted as $\mathbf{P}=[\mathbf{p}_1,\dots,\mathbf{p}_Q]$, $\mathbf{p}_q=[\alpha_{q,1}\left\|\mathbf{w}_q\right\|^2,\dots,\alpha_{q,K}\left\|\mathbf{w}_q\right\|^2]^T$, $p_k=\alpha_{q,k}\left\|\mathbf{w}_q\right\|^2$ and $P_0 = \left|\mathbf{h}_{q,k}^{H}\mathbf{G}{\mathbf{w}_q}\right|^{2}\alpha_{q,k}$ is the maximum total transmit power. Hence, the transmit power can be simplified as
\begin{equation}
    \label{transmitpower_sim}
    \sum_{q}^{Q}\left\|\mathbf{w}_q\right\|^2_2 = \text{Tr}(\mathbf{W}\mathbf{W}^H)=\text{Tr}({(\mathbf{U}^H(\mathbf{X}^*)\mathbf{U}(\mathbf{X}^*))}^{-1}{\mathbf{P}}).
\end{equation}

Substituting (\ref{ZFtransmitbeam}) into (\ref{SINR}), we can obtain the simplified SINR as follows
\begin{equation}
    \label{SINR_sim}
    {{\mathrm{SINR}}_{q,k}}= \frac{P_0}{P_0\sum\limits_{i=k+1}^{K}\alpha_{q,i} + \sigma^2},
\end{equation}
if UE $k$ is the strongest user in cluster $q$, the SINR can be further simplified as ${\mathrm{SINR}}_{q,k}^{\text{SIC}}=\frac{P_0}{\sigma^2}$ adopted SIC technology. Thus, $\mathbf{P}_0$ can be simplified as
\begin{subequations}
    \label{min_transmitpower5}
    \begin{align}
            \mathbf{P}_5: \quad & \min_{\mathbf{W},\mathbf{U},\alpha_{q,k}}\text{Tr}({(\mathbf{U}^H(\mathbf{X}^*)\mathbf{U}(\mathbf{X}^*))}^{-1}\mathbf{P}),
            \label{min_p}\\
            \mathrm{s.t.~} \quad & {\mathrm{SINR}}_{q,k}^{\text{SIC}} \ge {\mathrm{SINR}}_{q,k}^{\min}
            \label{PO_constraint}\\
            & \text{(\ref{rateconstraints})-(\ref{PA_x})}.
    \end{align}
\end{subequations}

Substituting to the optimal position $\mathbf{X}^*$, we can obtain the optimal power allocation $\alpha_{q,k}^*$ to minimize this objective function, that means the optimal power matrix $\mathbf{P}^*$. Finally, the optimal pinching beamforming matrix is derived as $\mathbf{U}^*=\mathbf{h}^{H}_{q,k}\mathbf{G}(\mathbf{X}^*)$ and the optimal transmit beamforming matrix is derived as $\mathbf{W}^*=\mathbf{U}(\mathbf{X}^*){(\mathbf{U}^H(\mathbf{X}^*)\mathbf{U}(\mathbf{X}^*))}^{-1}{\sqrt{\mathbf{P^*}}}$.
In summary, the joint PSO-based pinching antenna position optimization and ZF-based transmit and pinching beamforming and power allocation optimization algorithm is shown as Algorithm \ref{PSOZF}.

\begin{algorithm}[!ht]
    \caption{PSO-ZF-Based Joint Optimization of Transmit And Pinching Beamforming And Power Allocation Algorithm}
    \label{PSOZF}
    \begin{algorithmic}[1]
    \REQUIRE $Q$, $K$, $N$, $L$, $x^{\max}$, minimum spacing $\Delta$, fitness function $f_0{(x_{n,l})}=\sum_{Q} \| \mathbf{w}_q \|^2$, ${\mathrm{SINR}}_{q,k}^{\min}$, $\sigma^2$, population size $P$, $T$, PSO parameters $a_0$, $a_1$, $a_2$
    \STATE Initialize particle positions $x_{n,l}$ and velocities $v_{n,l}$
    \STATE Evaluate fitness $\mathbf{P}_0$ based on transmit power under SINR constraints
    \STATE Set local best position $J_{n,l} \gets x_{n,l}$, global best position $O_{n,l} \gets \arg\min f_0(J_{n,l})$
    \STATE {\textbf{STEP A: Pinching Beamforming (including PA Position) Optimization}}
    \FOR {$t = 1$ to $T$}
      \FOR {$n=1$ to $N$}
        \FOR{$l = 1$ to $L$}
            \STATE Update velocity
            $v_{n,l}$ = $a_0 v_{n,l} + a_1 r_1 (J_{n,l}-x_{n,l}) + a_2 r_2 (O_{n,l}-x_{n,l})$
            \STATE Update position $x_{n,l}=x_{n,l} + v_{n,l}$ with $x_{n,l} - x_{n,l-1} \ge \Delta$, $x_{n,l} \in [0, x^{\max}]$
            \STATE Compute channel $\tilde{h}(x_{n,l})$ using strongest users from each cluster, ${\mathrm{SINR}}_{q,k}$ for all users
            \IF{${\mathrm{SINR}_{q,k}} \geq {\mathrm{SINR}_{q,k}}^{\min}$, $\forall q,k$}
                \STATE Compute fitness function: $f_0(x_{n,l})$
            \ELSE
                \STATE Set $f_0(x_{n,l})$ by large penalty
            \ENDIF
            \IF{$f_0(x_{n,l}) < f_0(J_{n,l})$}
                \STATE Update local best position $J_{n,l} = x_{n,l}$
            \ENDIF
        \ENDFOR
      \ENDFOR
      \STATE Update global best position $O_{n,l} = \arg\min f_0(x_{n,l})$
    \ENDFOR
    \STATE {\textbf{STEP B: Transmit Beamforming Optimization}}
    \STATE Calculate ZF directions by (\ref{ZFtransmitbeam})
        \FOR{$q=1$ to $Q$}
            \FOR{$k=1$ to $K$}
            \STATE Compute optimal $\alpha_{q,k}$ from (\ref{rateconstraints})-(\ref{PA_x}),(\ref{PO_constraint}) constraints recursively
            \STATE Compute $f_0(\mathbf{X}^*)$ and $\mathbf{P} = \sum \alpha_{q,k} f_0(\mathbf{X}^*)$
            \ENDFOR
        \ENDFOR
        \STATE Compute optimal transmit beamforming $\mathbf{W}^*$, optimal pinching beamforming $\mathbf{U}(\mathbf{X}^*)$
    \ENSURE $\mathbf{X}^* \gets \left\{O_{n,l}\right\}$, $\mathbf{U}^*$, $\mathbf{W}^*$, $\alpha_{q,k}^*$, $\mathbf{P}^*$
    \end{algorithmic}
\end{algorithm}

\subsection{Computational Complexity Analysis}
From the above derivation, for subproblem $\mathbf{P}_0$, the PSO-based method is utilized to obtain the optimal position matrix PA, and the computational complexity of this is $\mathcal{O}(MK)$. For subproblem (\ref{min_transmitpower5}), during $Q$ iterations, matrix inversions need to be computed at each iteration based on ZF, the computational complexity is $\mathcal{O}(QK^2)$. Thus, the total computational complexity of the algorithm is $\mathcal{O}(MK+QK^2)$.

\section{Simulation Results}
In this section, we provide numerical results to demonstrate the effectiveness of the proposed methods. The parameters are set as follows, unless stated otherwise. In the NOMA assisted PASS framework, the BS serves $N=4$ wavegudies, which equipped with $L=4$ flexible PAs and fed by a dedicated RF chain, serving $K=2$ users in each cluster. The $Q=4$ clusters are near to the corresponding waveguides. And all users are on the ground and their heights are 0, their distribution ranges are from 0 to $x^{\max}=30$ meters along $x$-axis and from 0 to $y_N=12$ meters. Let the $S$ be the spatial range of users, which satisfies $x^{\max} \in [0, S]$. We assume that users in the cluster have the same $y$-axis positions as that of the corresponding waveguide. All PAs and waveguide have the same fixed heights $d_0^{\text{z}}=10 m$, the adjacent distance between two waveguides is $d_0^{\text{y}}=3$, the maximum length of each waveguide is $x^{\max}=30$ meters. Furthermore, maximum transmit power is set to $P_0 = 20$ dBm, the noise power is $\sigma^2=-80$ dBm, carrier frequency is $f_c= 15$ GHz, effective refractive index of dielectric waveguide $n_{\text{eff}}=1.4$, and the minimum SINR of user is set to 20 dB. For the MM-PDD method, the residual tolerance is set to $\epsilon=10^{-6}$, and the initial penalty is set to $\rho^{(0)}=10^{-4}$. For the PSO and ZF method, the swarm size is 30, the inertia weight is $a_0=0.7$, cognitive and social ratios are $a_1=1.5$ and $a_2=1.5$, respectively.

In the following, we consider the comparisons of NOMA assisted PASS with two methods, i.e., MM-PDD and PSO-ZF, and conventional MIMO-NOMA \cite{yu2016mmwave}. For simplification in the following analysis, the MIMO-NOMA represent the conventional NOMA assisted massive MIMO, MM-PDD and PSO-ZF represent MM-PDD-based and PSO-ZF-based methods in the proposed NOMA assisted PASS framework, respectively.

Aligned with the proposed NOMA assisted PASS framework, the conventional massive MIMO-NOMA system deploys a single BS at the origin, which is equipped with a uniform planar array (UPA) consisting of $N$ RF chains and $N\times L$ antennas. 
Each RF chain is connected to a subarray of $L$ antennas with half-wavelength antenna spacing through $L$ phase shifters to reduce the hardware costs. PDD algorithm is adopted to jointly optimize the analog beamforming, transmit beamforming, and power allocation for comparisons.

Fig. 2 demonstrates the convergence behaviors of the proposed PSO-ZF and MM-PDD algorithms. Fig. 2(a) shows convergence behavior of MM-PDD for NOMA assisted PASS. The transmit power decreases from 14 dBm to 11 dBm and converges after around 50 iterations. Fig. 2(b) illustrates the transmit power reduction of the PSO-ZF algorithm in each iteration under swarm size of 30. It can be observed that the power reduction per iteration is initially high (around 0.09dB), and then decreases rapidly as the algorithm converges. By the 21-$th$ iteration, the change of transmit power becomes negligible (0.00519814dB), which indicates the fast convergence and robustness of PSO-ZF. Fig. 2(c) further demonstrates the convergence of PSO-ZF with different swarm size (10,20,30). It can be observed that larger swarm sizes lead to lower final transmit power and faster convergence and small error range area. When swarm size is 30, the error variance (error range) is the smallest among these three configurations, which indicates the its stability. Increasing swarm size improves both the quality and robustness of the convergence, justifying the use of moderate-to-large populations, i.e., 30, for PSO-ZF in high-precision optimization. Compared to PSO-ZF, the MM-PDD method shows slower convergence, requires more iterations about 30, and tends to reach higher transmit power.

\begin{figure}[htbp]
    \centering
    \begin{subfloat}[]{\includegraphics[width=3.5in]{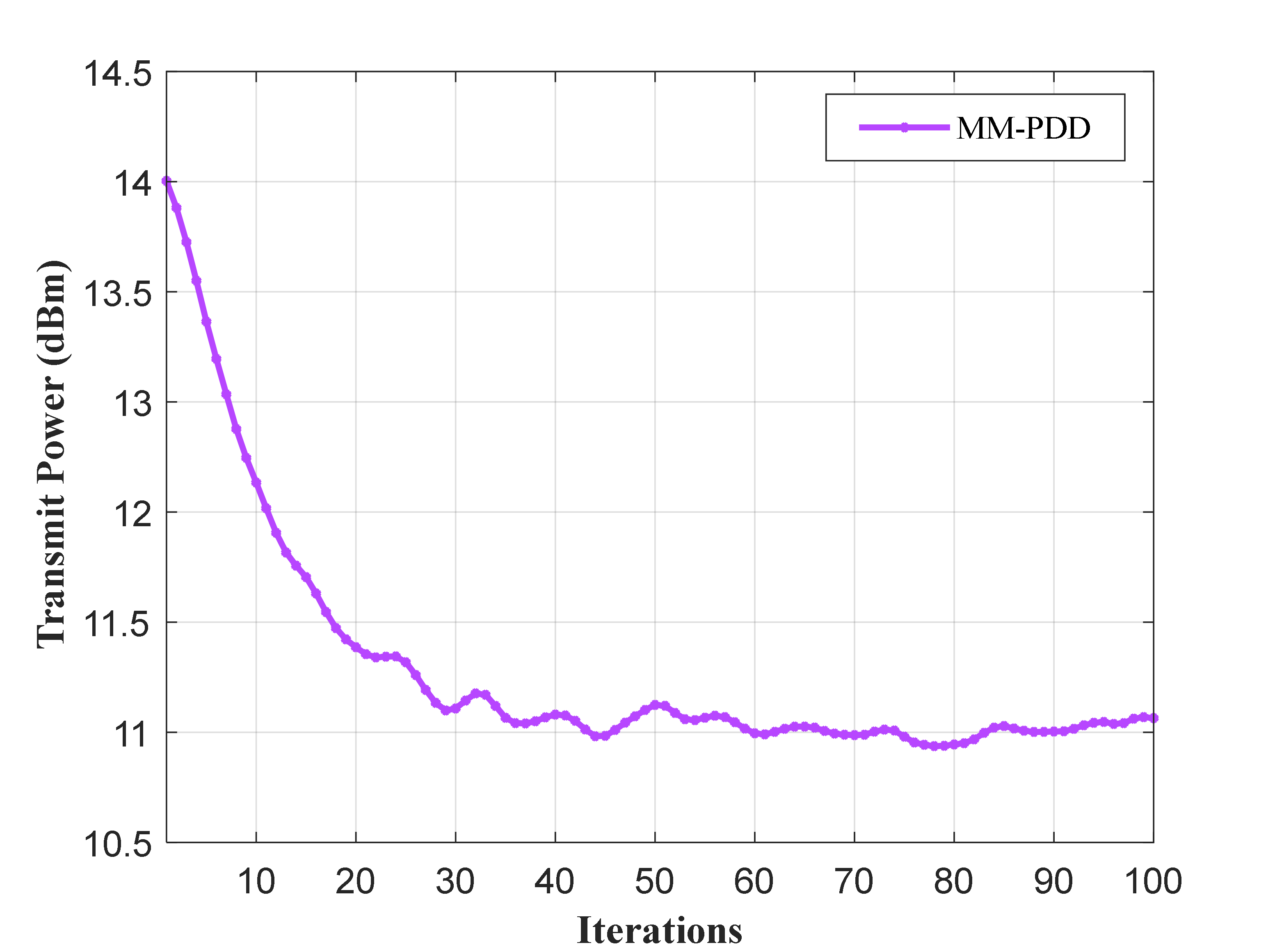}
        \label{fig_2a}}
    \end{subfloat}
    \centering
    \begin{subfloat}[]{\includegraphics[width=3.5in]{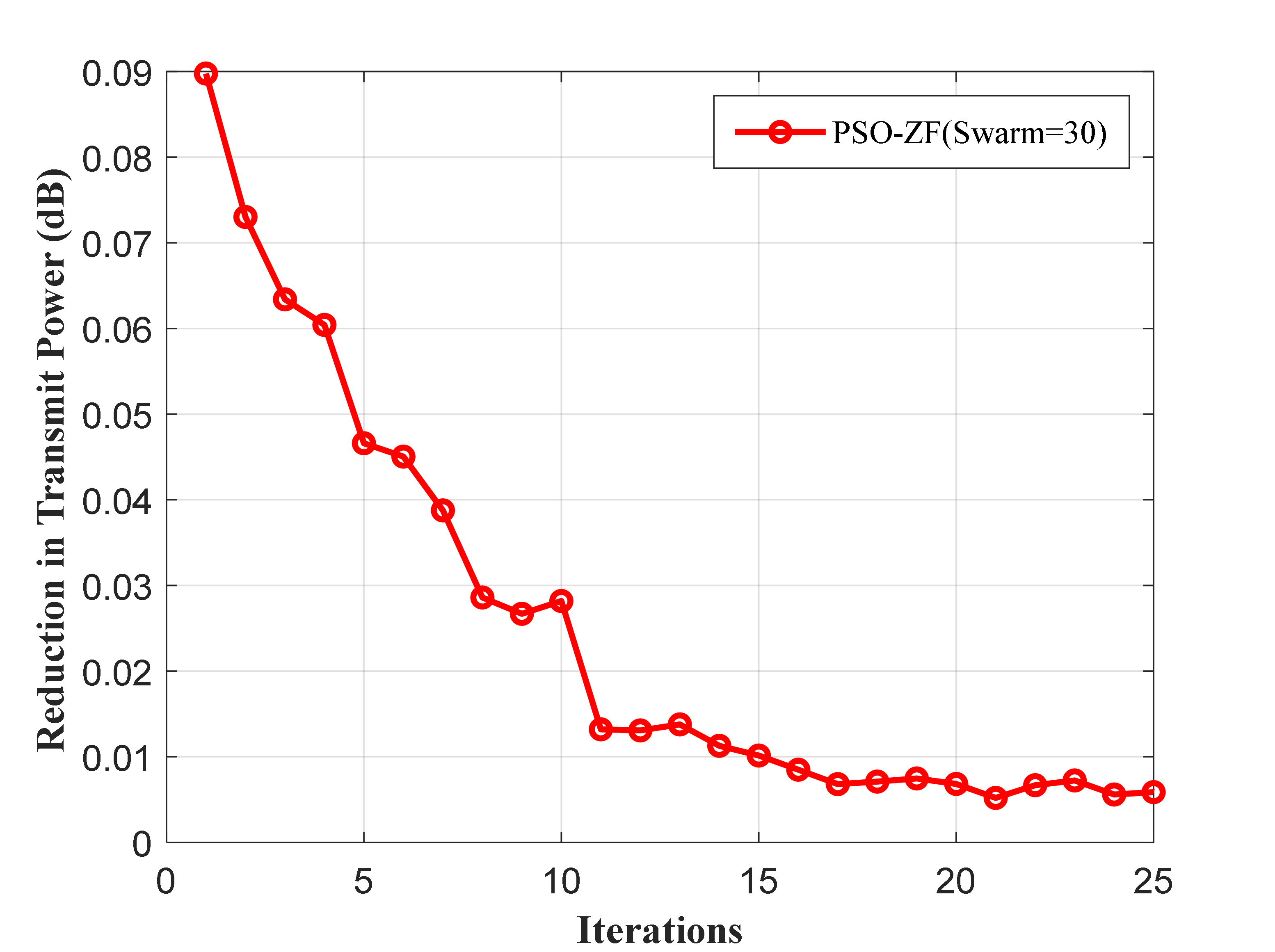}
    \label{fig_2b}}
    \end{subfloat}
    \centering
    \begin{subfloat}[]{\includegraphics[width=3.4in]{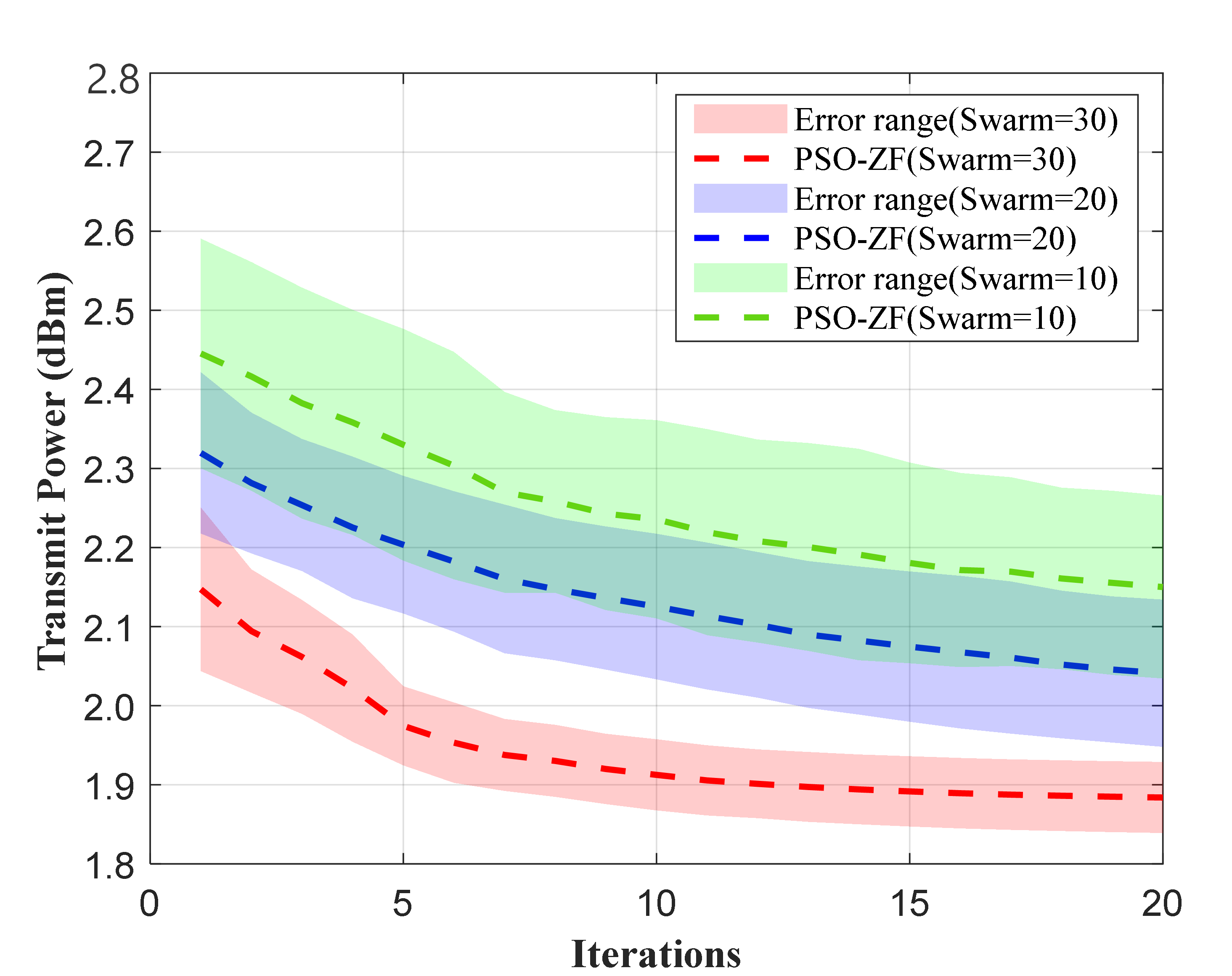}
    \label{fig_2c}}
    \end{subfloat}
    \caption{Comparisons of convergence behaviors of the developed algorithms. (a) convergence versus iterations of MM-PDD. (b) iteration-wise transmit power drop of PSO-ZF. (c) convergence versus iterations with three swarm sizes \{10,20,30\} of PSO-ZF.}
    \label{fig_2}
  \end{figure}




Fig. 3 demonstrates the impact of PA distance on the transmit power about PSO-ZF, MM-PDD and MIMO-NOMA. This evaluation is conducted for two different number of PAs $L=4$ and $L=8$ per waveguide. The transmit power increases as the distance of PA increases. Both MM-PDD and PSO-ZF of PASS consistently outperform the conventional MIMO-NOMA, with PSO-ZF showing the best performance. The gap between PASS and the conventional MIMO-NOMA widens especially at lower $S$. When the PA's distance is 30m and $L=4$, PSO-ZF decreases the total transmit power by 39.06$\%$ and 87.51$\%$, compared to MM-PDD and conventional massive MIMO-NOMA. This is because larger $S$ implies that the closest distance of PAs may reduces array gain and increasing path loss.

\begin{figure}[htbp]
    \centering
    \includegraphics[width=3.4in]{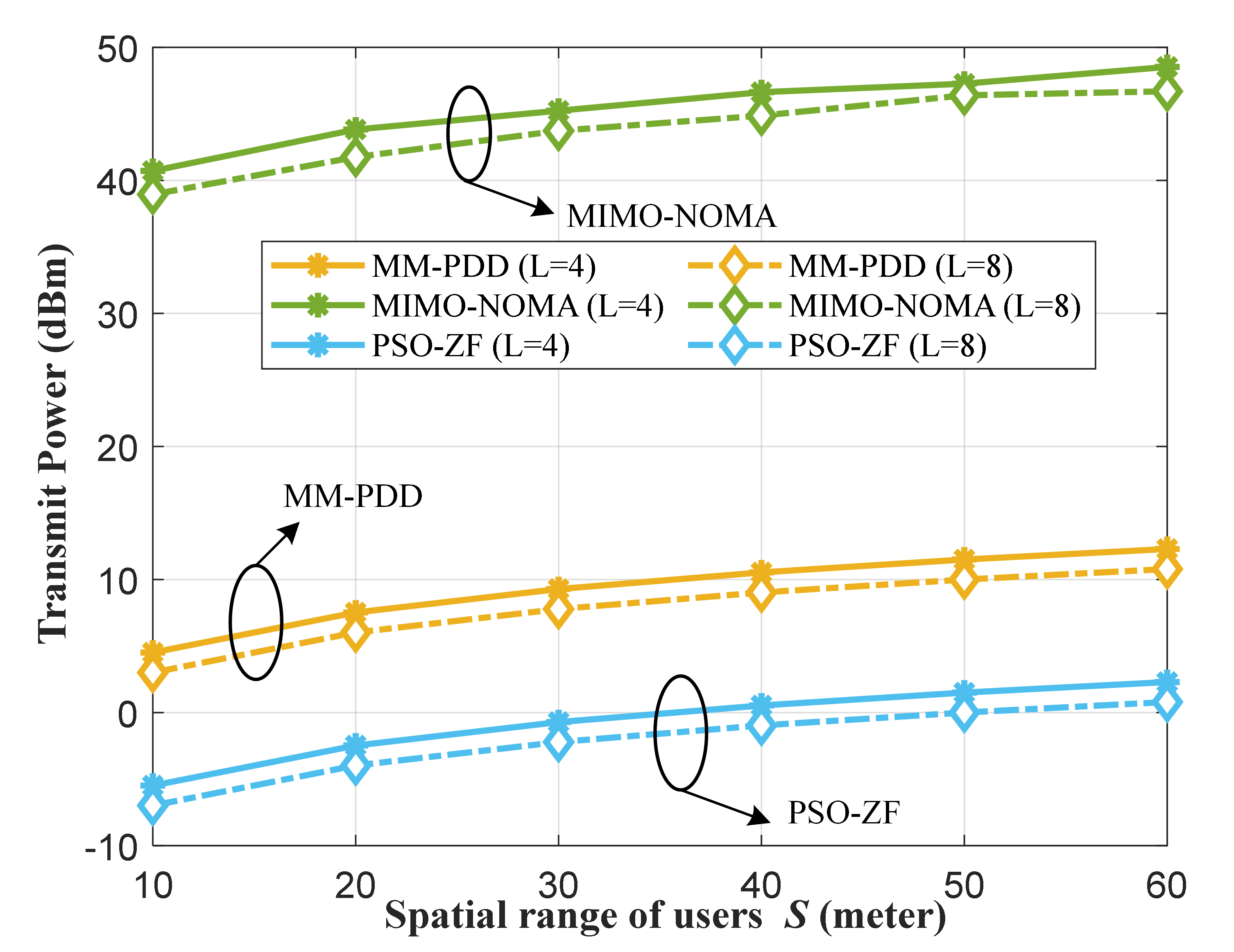}
    \caption{Comparisons of transmit power versus different distances of PAs.}
    \label{fig_3}
  \end{figure}

Fig. 4 presents the transmit power under varying numbers of PAs per waveguide $L\in \{2,4,6,8 \}$ with different rate requirements of users, for the proposed MM-PDD and PSO-ZF algorithms as well as the conventional MIMO-NOMA. For all schemes, increasing the number of pinching antennas per waveguide leads to reduced transmit power. With $L=4$ and 20bps/Hz minimum rate requirement, PSO-ZF decreases the transmit power by 32.55$\%$ and 96.37$\%$ compared to MM-PDD and conventional MIMO-NOMA. And MM-PDD reduces the transmit power by 60.02$\%$ compared to coventional MIMO-NOMA. Across all PA configurations $L\in \left\{2,4,6,8\right\}$ and rate requirements, PSO-ZF achieves the lowest transmit power, demonstrating the superiority of PASS in leveraging beamforming and antenna placement flexibility.

\begin{figure}[htbp]
    \centering
    \includegraphics[width=3.4in]{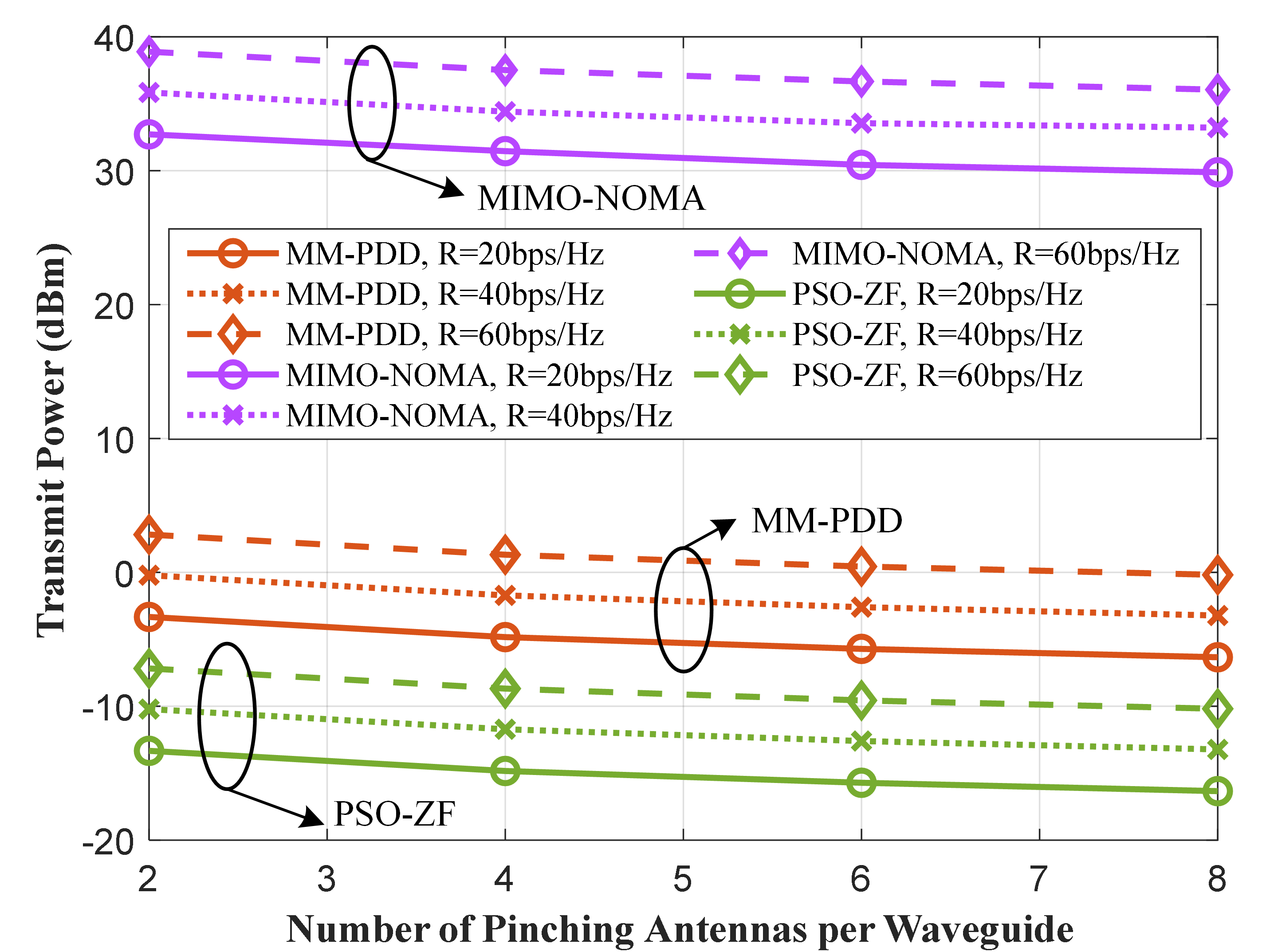}
    \caption{Comparisons of transmit power versus different dynamic numbers of PAs per waveguide $L\in \{2,4,6,8\}$.}
    \label{fig_4}
\end{figure}

\begin{figure}[htbp]
    \centering
    \includegraphics[width=3.4in]{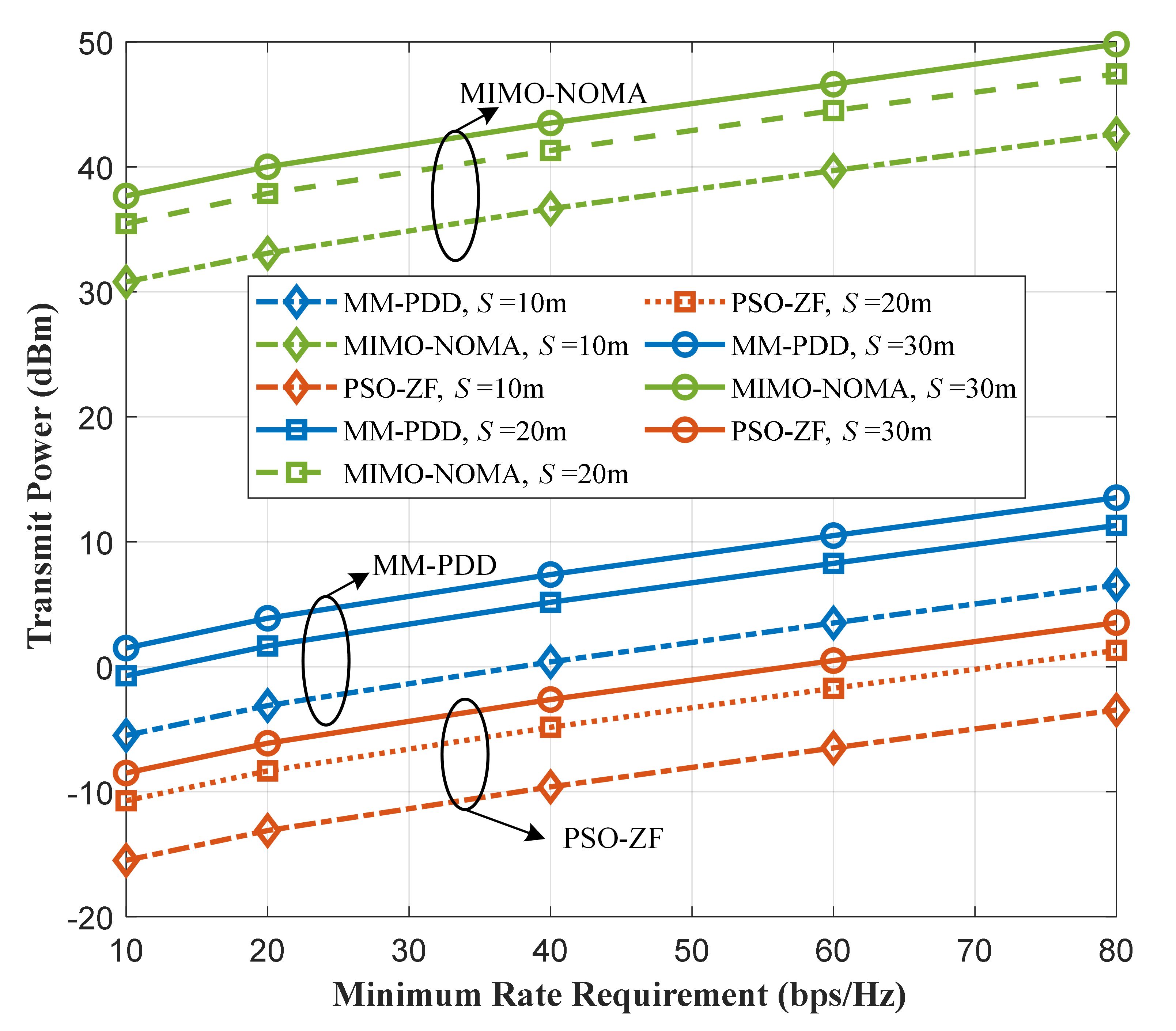}
    \caption{Comparisons of transmit power versus different rate requirements.}
    \label{fig_5}
\end{figure}

Fig. 5 compares the transmit power of different algorithms under different rate requirements of users. As expected, the minimum rate requiremet of user increase, the corresponding total transmit power becomes higher. When the minimum rate requiremet is 40bps/Hz and $S=$30m, PSO-ZF decreases the transmit power by 27.39$\%$ and 95.22$\%$, compared to MM-PDD and conventional MIMO-NOMA. This validates the ability of PSO-ZF algorithm to jointly optimize pinching positions and beamforming to minimize power consumption, especially under high data rate demands. MM-PDD reduces the transmit power by 86.01$\%$, compared to conventional MIMO-NOMA. Across all rate and distance configurations of PAs, NOMA assisted PASS consistently achieves the lowest transmit power, often outperforming conventional MIMO-NOMA by a large margin.

\begin{figure}[htbp]
    \centering
    \includegraphics[width=3.4in]{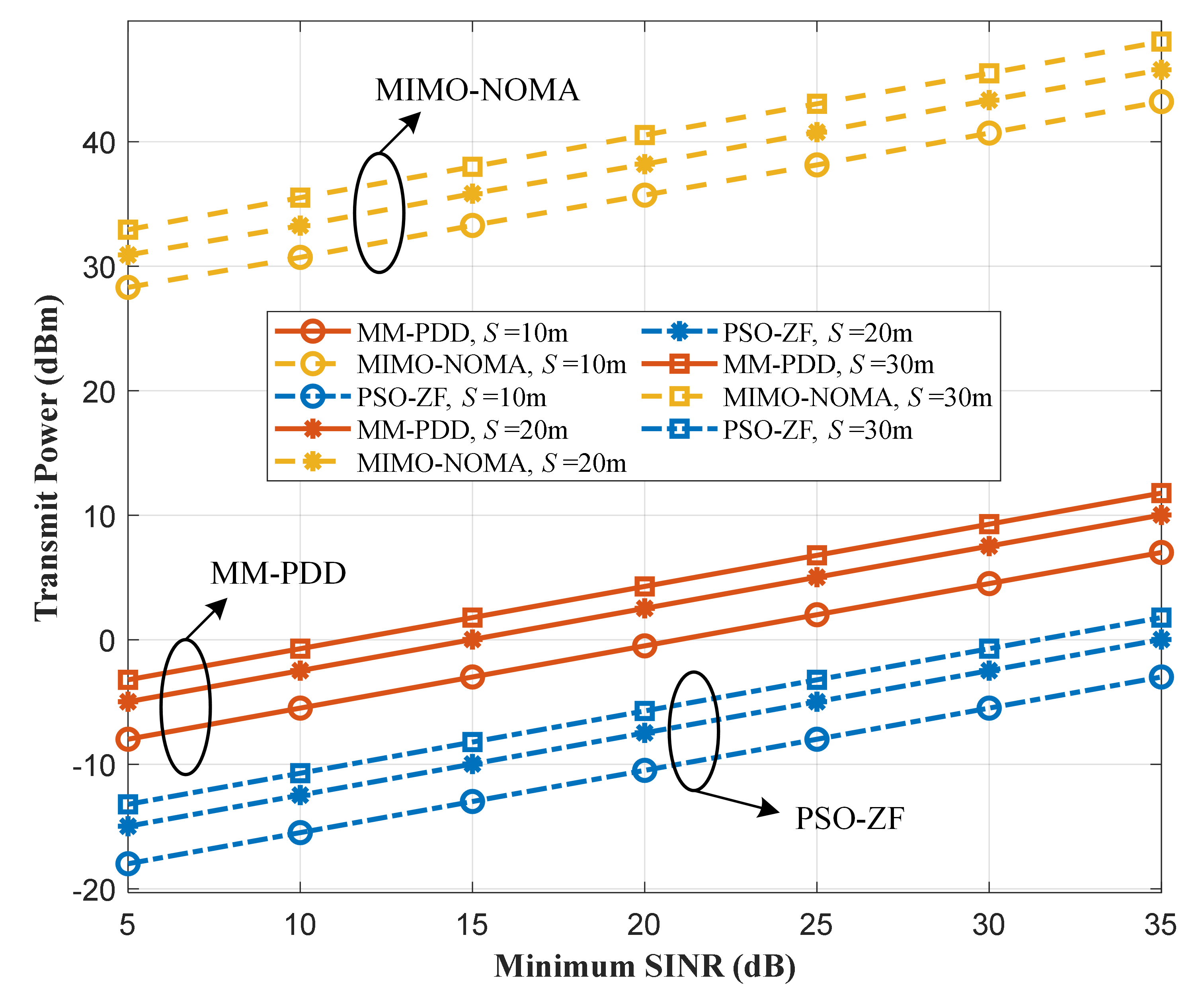}
    \caption{Comparisons of transmit power versus different minimum SINR.}
    \label{fig_6}
  \end{figure}

Fig. 6 illustrates the performance of the proposed PSO-ZF and MM-PDD algorithms and the benchmark MIMO-NOMA algorithm in terms of the transmit power, under varying minimum SINR requiremnets from 5dB to 35dB. We can see that the transmit power increases with the higher minimum SINR requirement.
Both MM-PDD and PSO-ZF algorithms achieve significantly lower transmit power than conventional MIMO-NOMA under all SINR and configurations. When the minimum SINR of users is 20dB and $S=30$m, PSO-ZF significantly reduces the transmit power by 95.32$\%$ compared to the conventional MIMO-NOMA. This confirms the strong capability of PASS to combat free-space pathloss by enabling near-user antenna deployment and effective LoS channel construction. PSO-ZF achieves 35.16$\%$ transmit power reduction compared to MM-PDD at the 20dB minimum SINR.

\section{Conclusion}
In this paper, a novel NOMA assisted PASS framework has been proposed for downlink multi-user MIMO communications, which deploys dielectric waveguides to adjust the positions of PAs to configure the large-scale path loss and radiated signal.
The transmit power minimization problem has been formulated, which jointly optimizes the transmit and pinching beamforming and power allocation. Both gradient-based and swarm-based methods have been developed. For gradient-based method, we first converted the resulting nonconvex coupling problem into a tractable form. Then, a MM-PDD algorithm has been proposed to obtain the stationary closed-form solutions. For swarm-based method, we have explored the PSO-ZF algorithm to minimize the transmit power. Specifically, the PSO method has been utilized to build position-seeking particles for pinching beamforming, where ZF-based transmit beamforming has been adopted by each particle for fast computation. The simulation results have demonstrated the effectiveness of the proposed NOMA assisted PASS framework over conventional NOMA assisted massive MIMO systems. Compared to conventional massive MIMO-NOMA, the proposed framework can decrease the transmit power by 95.22$\%$. Furthermore, compared to MM-PDD algorithm, the PSO-ZF algorithm can avoid stuck in undesired local optimum with an acceptable computational complexity.

\bibliographystyle{IEEEtran}
\def\baselinestretch{1}
\bibliography{IEEEabrv,NOMA_PASS}


\end{document}